\documentclass[aps,superscriptaddress,nofootinbib,a4paper,11pt]{revtex4-2}
\usepackage{amsfonts,amsmath,amssymb}
\usepackage{graphicx}
\usepackage{natbib}
\usepackage{bm}
\usepackage{slashed}
\usepackage[usenames,dvipsnames]{xcolor}

\definecolor{venetianred}{rgb}{0.78, 0.03, 0.08}
\definecolor{midnightblue}{rgb}{0.1, 0.1, 0.44}
\definecolor{regalia}{rgb}{0.32, 0.18, 0.5}

\usepackage{hyperref} 
\hypersetup{
    colorlinks,
    linkcolor={venetianred},
    citecolor={midnightblue},
    urlcolor={regalia}
}

\DeclareMathOperator*{\sumint}{%
\mathchoice%
  {\ooalign{$\displaystyle\sum$\cr\hidewidth$\displaystyle\int$\hidewidth\cr}}
  {\ooalign{\raisebox{.14\height}{\scalebox{.7}{$\textstyle\sum$}}\cr\hidewidth$\textstyle\int$\hidewidth\cr}}
  {\ooalign{\raisebox{.2\height}{\scalebox{.6}{$\scriptstyle\sum$}}\cr$\scriptstyle\int$\cr}}
  {\ooalign{\raisebox{.2\height}{\scalebox{.6}{$\scriptstyle\sum$}}\cr$\scriptstyle\int$\cr}}
}

\def\sm[#1]{{\scalebox{.7}{$\scriptscriptstyle #1$}}}
\def\sv[#1]{{\scalebox{.9}{$\scriptscriptstyle #1$}}}
\def\sl[#1]{{\scalebox{1.1}{$\scriptscriptstyle #1$}}}

\def\trD{\mathrm{Tr}_\sl[\rm D]}
\def\per{\!\sl[\perp]}
\makeatletter
\newcommand{\npar}{\mathrel{\mathpalette\new@parallel\relax}}
\newcommand{\new@parallel}[2]{%
  \begingroup
  \sbox\z@{$#1T$}
  \resizebox{!}{\ht\z@}{\raisebox{\depth}{$\m@th#1 ||$}}%
  \endgroup
}
\makeatother

\begin{document}

\title{\sc\Large{Thermomagnetic effects on light pseudo-scalar meson masses within the SU(3) Nambu-Jona--Lasinio model} \vspace*{5mm}}

\author{M\'aximo Coppola\,}\email{maximocoppola@cnea.gob.ar}
\affiliation{Departamento de F\'isica Te\'orica, Comisi\'on Nacional de Energ\'ia At\'omica,
Av. Libertador 8250, (1429) Buenos Aires, Argentina}
\author{William R. Tavares\,} \email{tavares.william@ce.uerj.br}
\affiliation{Departamento de F\'isica Te\'orica, Universidade do Estado do Rio de Janeiro,
20550-013 Rio de Janeiro, Rio de Janeiro, Brazil}

\author{Sidney S. Avancini\,} \email{sidney.avancini@ufsc.br}
\affiliation{Departamento de F\'isica, Universidade Federal de Santa Catarina, 88040-900 Florian\'opolis, Santa Catarina, Brazil}
\author{Joana C. Sodr\'e\,} \email{joana.sodre@gmail.com}
\affiliation{Departamento de F\'isica, Universidade Federal de Santa Catarina, 88040-900 Florian\'opolis, Santa Catarina, Brazil}
\author{Norberto N. Scoccola\, }\email{norbertoscoccola@cnea.gob.ar}
\affiliation{Departamento de F\'isica Te\'orica, Comisi\'on Nacional de Energ\'ia At\'omica,
Av. Libertador 8250, (1429) Buenos Aires, Argentina}
\affiliation{CONICET, Rivadavia 1917, (1033) Buenos Aires, Argentina \vspace*{1.5cm}}


\begin{abstract}
{\centering \large Abstract \par}

We calculate the screening masses of pseudoscalar mesons in a hot and strongly magnetized medium within the framework of the SU(3) Nambu-Jona--Lasinio model, using a magnetic field-independent regularization scheme. 
Inverse magnetic catalysis (IMC) is implemented through the use of a magnetic field-dependent coupling $G(B)$, fitted to reproduce lattice quantum chromodynamics (QCD) results for the pseudocritical chiral transition temperature $T_c^B$. 
For the external homogeneous magnetic field considered, neutral screening masses separate in two types: perpendicular and parallel to the direction of the field, while for charged mesons only  parallel energies can be defined for each Landau level. 
We obtain $m_\sl[\mathrm{scr},{\per}] > m_\sl[\mathrm{scr},{\npar}]$, as expected from causality.
Thermally, all screening energies are almost constant until some critical temperature, whose behavior is correlated with $T_c^B$.
They rapidly increase around this value, keeping a steady enhancement afterward due to thermal excitation. 
Magnetically, neutral parallel masses are enhanced (suppressed) at high temperatures when considering $G(B)$ ($G$).
Perpendicular ones display a non-monotonic magnetic behavior for $G$ (due to increasing $T_c^B$) when $T \lesssim 500$~MeV, but become magnetically enhanced when $T \gtrsim 500$~MeV. For $G(B)$ they always increase with $B$.
Charged parallel energies are always magnetically enhanced, for both couplings.
In the high-temperature limit, we show that both neutral and charged screening energies converge to $2\pi T$.
At $B=0$ the model overestimates the remaining quark interaction in this regime.
At $B\neq 0$ we find that, when IMC is accounted for, the interaction is suppressed as $B$ increases, a fact that appears to be at odds with currently available lattice QCD results.

\end{abstract}

\maketitle
\clearpage

\section{Introduction}

The dynamical breaking of chiral symmetry in the vacuum of quantum chromodynamics (QCD) has been one of the most enlightening concepts of modern physics. 
The mechanism is amplified by the presence of external magnetic fields, modifying its structure and inducing new exciting phenomena. 
Magnetic effects on QCD matter are expected to be relevant for diverse physical scenarios where powerful magnetic fields can be present, such as peripheral heavy ion collisions, the electroweak phase transition of the primordial universe~\cite{Grasso:2000wj} and the interior of certain compact objects known as magnetars~\cite{Duncan:1992hi}.
In some of these cases, it has been predicted that such fields can reach values as high as 
$eB\sim 10^{20}$~G~\cite{Ferrer:2010,Deng:2012,Tuchin:2013ie}. 
Since these fields are stronger than the (squared) QCD scale, they profoundly affect the properties of strongly interacting matter~\cite{Kharzeev2012,Andersen:2014xxa,Miransky:2015ava,Endrodi:2024cqn}. 
The most prominent example is the altering of the nature of chiral phase transitions~\cite{Andersen:2014xxa,Huang:2015oca,DElia:2022}.
Moreover, because of the interplay with quantum anomalies, these strong fields also give rise to a series of intricate transport phenomena, causing numerous phenomenological consequences~\cite{Miransky:2015ava} such as the chiral separation effect~\cite{Gorbar:2013upa}, chiral magnetic waves~\cite{Kharzeev:2010gd,Burnier:2011bf,Yee:2013cya,STAR:2015wza,Shovkovy:2018tks} and the chiral magnetic effect (CME)~\cite{Fukushima:2012vr,Kharzeev:2014,Li:2020dwr}, to name a few.
In particular, CME observables are used in experiments to aid in the determination of the intensity of such strong fields~\cite{Huang:2015oca}.

Magnetic fields are not the only parameters relevant for these scenarios: the presence of temperature, density and other parameters should also be taken into account.
In particular, heavy ion collisions are hot environments, where quark-gluon plasmas (QGP) have been discovered to be formed~\cite{Nouicer:2015jrf}.
The QGP corresponds to a chirally restored regime where quarks are not confined but still strongly coupled.
Lattice QCD (LQCD) simulations have found that the chiral transition is expected to be a smooth crossover, associated with a pseudocritical value of $T_c^0|_{\rm LQCD}=156.5$~MeV~\cite{HotQCD:2018pds}.

From the theoretical side, many efforts have been made to understand the impact that external magnetic fields have on the chiral symmetry breaking mechanism.
One notable achievement is the well known magnetic catalysis (MC) effect~\cite{DElia:2010abb}, corresponding to the enhancement of the chiral condensate by the magnetic field at low temperatures~\cite{Shovkovy2013,Miransky:2015ava}. 
On the other hand, close to the pseudocritical temperature we have the interesting phenomenon of inverse magnetic catalysis (IMC), defined by a non-monotonic behavior of the chiral condensate for strong enough values of magnetic fields, i.e., $eB\sim 0.2$~GeV$^2$~\cite{Bali:2011qj,Bali:2012zg}.
Closely related is the ``deconfinement catalysis'' effect, which refers to the decrease of the pseudocritical temperature with the magnetic field~\cite{Bali:2011qj,DElia:2018xwo,Endrodi:2019zrl}.
MC is well established by theoretical descriptions in the literature~\cite{Andersen:2014xxa,Miransky:2015ava}. 
However, the IMC effect, predicted by LQCD~\cite{Bali:2011qj,Bali:2012zg,Bali2014,Bornyakov2014}, is not fully understood in mean field effective models and several different approaches have been explored in order to reproduce IMC in effective descriptions~\cite{Andersen:2014xxa,Mao:2016fha,Dumm:2017,Bandyopadhyay:2020zte,Andersen:2021lnk,Ayala:2021nhx}. 

In addition to the impact of the magnetic field on the properties of the QCD phase diagram, it is also important to study its effect on the excitations arising from hadronization processes. 
There is recent literature about low lying mesons in strong magnetic fields by 
LQCD~\cite{Hidaka:2012mz,Luschevskaya:2012xd,Luschevskaya:2015bea,Andreichikov:2016ayj,Bali:2017ian,Ding:2021} and 
effective model descriptions, such as the Nambu--Jona-Lasinio model and its extensions~\cite{Avancini:2016fgq,Avancini:2018svs,Avancini:2021pmi,Carlomagno:2022arc,Carlomagno:2022inu,Coppola:2023mmq}, linear sigma models coupled with quarks~\cite{Ayala:2018zat,Das2020,Ayala:2020dxs,Das:2022mic,Ayala:2023llp}, 
effective chiral lagrangians~\cite{Agasian:2001ym,Andersen:2012zc,Andreichikov:2018wrc}, 
$\bar{q}-q$ formalism~\cite{Orlovsky:2013gha,Simonov:2015xta}, 
QCD sum rules~\cite{Dominguez:2018njv} and 
non-relativistic approaches~\cite{Hattori:2015aki,Kojo:2021gvm}.
Particularly, the NJL model and its extensions has been extensively explored at zero temperature, showing qualitatively good agreement with LQCD, for example, for the $\pi^0$ pole mass~\cite{Avancini:2015ady,Avancini:2017gck,Coppola:2018vkw,Coppola:2019uyr,Dumm:2020muy,Avancini:2021pmi}, which displays a monotonic decrease with the magnetic field. 
For $\pi^{\pm}$, the NJL model predicts intriguing results when compared with different lattice groups.
At small magnetic field strengths, both approaches agree on an initial enhancement of the energy~\cite{Coppola:2018vkw,Coppola:2019uyr}.
Then, for $eB \gtrsim 0.6$~GeV$^2$ the NJL model predicts a continuum steady increase, while some LQCD results show a non-monotonic behavior~\cite{Ding:2021}. 
For vector mesons, new inspiring results indicate good agreement for neutral~\cite{Avancini:2022qcp,Carlomagno:2022arc} and charged~\cite{Carlomagno:2022inu} $\rho$ mesons with different spin projections as compared with LQCD results~\cite{Hidaka:2012mz,Andreichikov:2016ayj,Bali:2017ian}, showing no evidence of $\rho^{\pm}$ condensation~\cite{Chernodub:2011,Liu:2014uwa}.
In fact more robust results are obtained when considering the mixing with the axial sector~\cite{Coppola:2023mmq}, leading to even better agreement with LQCD curves for both pions and $\rho$ mesons.
For simplicity, we do not include the mixing with the axial and vector sector in this work.
Nevertheless, their effect can be qualitatively mimicked by introducing a magnetic-field dependence for the coupling constants of the model, which we will consider in the numerical section.

Turning the attention to the hot magnetic medium, results for meson masses are still incipient and not widely investigated. 
LQCD simulations using (2+1)-flavors with highly improved staggered fermions have recently obtained results for the screening masses of pseudoscalar neutral mesons ($\pi^0$, $K^0$ and $\eta^0_{s\bar{s}}$), extracted from two-point spatial correlation functions~\cite{Ding:2022tqn}. 
Results from other approaches include the functional renormalization group~\cite{Kamikado:2013pya}, chiral perturbation theory~\cite{Colucci:2013zoa} and the NJL model, where there has been significant advances in the calculation of the curvature, pole and screening masses for neutral mesons within the 
two-flavor~\cite{Avancini:2018svs,Ghosh:2020qvg,Sheng:2020hge,Sheng:2021evj,Abreu:2021btt} and three-flavor~\cite{Mei:2022dkd} formulations, 
as well as with the non-local two-flavor version of the model~\cite{Dumm:2020muy}. 
More complete works with the inclusion of charged mesons with two flavors can be found in Refs.~\cite{Liu:2018zag,Wang:2018mpm,Li:2023rsy}. 

We recall here that both temperature and magnetic fields break the euclidean SO(4) Lorentz invariance, leading to a splitting of different types of meson masses as compared to the usual $q_1^2+q_2^2+q_3^2+q_4^2=m^2$ isotropic dispersion relation.
Temperature breaks the spatial-temporal symmetry related to boosts, symbolically written as 123-4, thus differentiating between pole (4) and screening (123) masses, which are defined as the decay constants associated with the temporal and spatial exponential decay of the mesonic correlation function at large distances, respectively.
Additionally, an homogeneous magnetic field aligned in the 3-direction (which will be considered in this work) breaks rotational symmetry between the field-direction and its perpendicular plane, i.e. 12-3, further splitting the screening masses between perpendicular (12) and parallel (3) types.
Because of asymptotic freedom, in the large temperature limit the mesonic system should consist essentially of two free quarks (at zero Matsubara mode). Thus, the screening mass is expected to be asymptotically proportional to $2\pi T$~\cite{Eletsky:1988an,Florkowski:1993bq}.
Notice that, in LQCD calculations, it is easier to calculate the screening mass than the pole mass, especially at large temperatures where the temporal extent of the lattice is associated with the temperature.

In this work we explore the effects of temperature and strong magnetic fields on the screening masses of the pseudoscalar nonet, in the context of a three-flavor Nambu--Jona-Lasinio model. 
The Kobayashi-Maskawa-'t Hooft (KMT) interaction is included to reproduce the breaking of the $U_A(1)$ symmetry, lifting the degeneracy between $\eta$ and $\eta'$.
While quark condensates emerge from the usual mean field approximation, mesons are treated as excitations whose masses are obtained by considering second order corrections in the bosonized Euclidean action of the model. 
The quark propagator in a constant magnetic field, considered in this work in the Schwinger form~\cite{Andersen:2014xxa,Schwinger:1951nm,Miransky:2015ava}, is used to evaluate meson polarization functions, which are diagonalized through different bases; Fourier basis for neutral mesons, and Ritus basis for charged ones~\cite{Ritus:1978cj,Coppola:2018vkw}. 
We adopt two different physical scenarios: without and with IMC. 
The first case is naturally reproduced in the model by mean field results, while for the second case we implement a magnetic field-dependence on the coupling of the four-point interaction channel, which we fix to reproduce the LQCD behavior of the pseudocritical temperature as a function of the magnetic field~\cite{Ferreira:2014kpa}. 
Temperature is included through the imaginary-time formalism, where the temporal momentum is written in terms of Matsubara frequencies~\cite{bailin1993introduction}. 
For the model regularization procedure, we adopt the magnetic field independent regularization (MFIR)~\cite{Ebert:2003,Allen:2015paa} scheme, which fully separates the vacuum, thermal and thermomagnetic contributions in our expressions and avoids usual unphysical results present in non-MFIR methods~\cite{Avancini:2019wed}.
This work is organized as follows: in Sec.~\ref{Formalism} the model and meson screening mass calculation formalism are briefly reviewed. 
Next, in Sec.~\ref{Numerical_Results} our numerical results are discussed.
We conclude and discuss future perspectives in Sec.~\ref{Conclusions}. 
Further details of the formalism are presented in Appendices \ref{app-A}--\ref{app-D}.


\vspace*{6mm}
\section{Theoretical formalism}\label{Formalism}

In this section, the main steps necessary for the calculation of screening masses in a hot magnetized medium are discussed. 
We aim to leave the text minimally self-contained: further details at zero temperature are given in~\cite{Avancini:2021pmi} and references therein.

\vspace*{2mm}
\subsection{Effective Lagrangian and mean field properties}
\label{sec2A}

The effective action of the three-flavor Nambu--Jona-Lasinio model (SU(3) NJL) in Euclidean space is given by
\begin{align}
S_{\rm E} & \:=\: \int d^4x  \left[ \bar{\psi}\left( -i\ \rlap/\!D  + \hat{m} \right)\psi 
+ {\cal L}_{\rm sym} + {\cal L}_{\rm det} \right] \ , \\[2mm]
{\cal L}_{\rm sym} &\:=\:  - G \sum_{a=0}^{8}\left[ \left( \bar{\psi}\lambda_{a}\psi \right)^{2}
    +\left( \bar{\psi}i\gamma_{5}\lambda_{a}\psi \right)^{2} \right] \ , \\[2mm]
{\cal L}_{\rm det} & \:=\: K \left( \det\left[ \bar{\psi}\left(1 + \gamma_{5} \right)\psi \right] +
\det\left[ \bar{\psi}\left(1 - \gamma_{5} \right)\psi \right] \right) \ ,
\end{align}
where $G$ and $K$ are coupling constants, $\psi=\left(\psi_u,\psi_d,\psi_s\right)^{T}$ represents a quark field with three
flavors and $\hat{m}=\mathrm{diag}\left( m_{u},m_{d},m_{s} \right)$ is the corresponding current quark mass matrix. 
In addition, $\lambda_{0}=\sqrt{2/3}\,I$, where $I$ is the unit matrix in three-flavor space, and $\lambda_{a}$ with $a=1,...,8$ denote the Gell-Mann matrices.
The action contains a U(3)$_L\times$ U(3)$_R$ symmetric four-point interaction term ${\cal L}_{\rm sym}$ and the determinantal six-point KMT interaction ${\cal L}_{\rm det}$ which breaks the $U_A(1)$ symmetry. 

The coupling of quarks to the external electromagnetic field ${\cal A}_\mu$ is implemented through the covariant derivative $D_{\mu}=\partial_\mu - i \hat Q {\cal A}_{\mu}$, where  $\hat Q=\mathrm{diag}\left( Q_{u},Q_{d},Q_{s} \right)$ represents the quark electric charge matrix with  $Q_u/2 = -Q_d = - Q_s = e/3$, $e$ being the proton electric charge. 
In the present work, we consider a static and constant magnetic field in the $3$-direction. 

For the calculation of meson masses we employ a bosonization procedure which we summarize below (see also Ref.~\cite{Avancini:2021pmi}).
As usual, we start with the partition function
\begin{align}
\mathcal{Z} \:=\: \int D\bar{\psi} D\psi \ e^{-S_\mathrm{E}} \ .
\end{align}
By using the functional delta functions
\begin{align}
\delta \left(s_a(x) - \bar{\psi}\lambda_{a}\psi \right) & \:=\: 
\int {\cal D} \sigma_a \; e^{\, \int \! d^4x \: \sigma_a(x) \left[ s_a(x) - \bar{\psi}\lambda_{a}\psi \right] }
\ , \nonumber \\[2mm]
\delta \left(p_a(x) - \bar{\psi}i\gamma_5 \lambda_{a}\psi \right) & \:=\: 
\int {\cal D} \pi_a \; e^{\, \int \! d^4x \: \pi_a(x) \left[ p_a(x) - \bar{\psi}i\gamma_{5}\lambda_{a}\psi \right] } \ ,
\end{align}
one can introduce scalar ${s}_a(x)$ and pseudoscalar ${p}_a(x)$ auxiliary fields together with their corresponding scalar $\sigma_a(x)$ and pseudoscalar $\pi_a(x)$ meson fields.
This way, the scalar $(\bar{\psi}\lambda_{a}\psi$) and pseudoscalar ($\bar{\psi}i\gamma_{5}\lambda_{a}\psi$) terms present in $S_{\rm E}$ are replaced by ${s}_a(x)$ and ${p}_a(x)$. 
The functional integration over the auxiliary variables ${s}_a(x)$ and ${p}_a(x)$  is performed through the stationary phase approximation (SPA) method, also known as saddle-point approximation~\cite{Reinhardt:1988xu,Ebert:1994mf,Osipov:2005sp}. 
The latter consists of choosing {$\tilde{\mbox{s}}_a(x)$ and $\tilde{\mbox{p}}_a(x)$} so as to minimize the term present in the exponential of the partition function integrand, disregarding second order terms. 
The calculation of the minimum yields a set of coupled equations where $\tilde{\mbox{s}}_a(x)$ and $\tilde{\mbox{p}}_a(x)$ have to be considered implicit functions of $\sigma_a(x)$ and $\pi_a(x)$
\begin{align}
& \sigma_a(x)  + 2 G \, \tilde{\mbox{s}}_a(x) - \dfrac{3}{4}K A_{abc}
\left[ \tilde{\mbox{s}}_b(x)  \; \tilde{\mbox{s}}_c(x)  - 
\tilde{\mbox{p}}_b(x)\; \tilde{\mbox{p}}_c(x) \right] \:=\: 0
\ ,\nonumber \\[3mm]
& \pi_a(x) + 2 G \, \tilde{\mbox{p}}_a(x) + 
\dfrac{3}{2} K A_{abc}\;  \tilde{\mbox{s}}_b(x)\; \tilde{\mbox{p}}_c(x) \:=\: 0 \ ,
\label{SPA}
\end{align}
where $A_{abc}$ stands for the totally symmetric SU(3) structure constants.
The last step is to use the mean-field approximation (MFA), where the bosonized action is expanded in powers of field fluctuations around the corresponding translational invariant mean field values $\bar{\sigma}_a$ and $\bar{\pi}_a$, namely
$\sigma_a(x)=\bar{\sigma}_a+\delta\sigma_a(x)$ and $\pi_a(x)=\bar{\pi}_a+\delta\pi_a(x)$.
Since charge conservation constrains the charge matrix to commute with the SU(3) generators, i.e. $[Q,\lambda_a]=0$, only diagonals components are different from zero. 
For the scalar sector we introduce $\bar{\sigma}=\lambda_0\bar{\sigma}_0+\lambda_3\bar{\sigma}_3+\lambda_8\bar{\sigma}_8 \equiv \mathrm{diag}(\bar{\sigma}_u,\bar{\sigma}_d,\bar{\sigma}_s)$, while expectation values of pseudoscalar boson fields are zero so as to respect the parity symmetry of the vacuum, $\bar{\pi}_a=0$.
Finally, one obtains the Euclidean action in the mean field approximation per unit of volume
\begin{align}
\dfrac{\bar{S}_\sl[\rm E]^\sl[\,\rm bos]}{V^{(4)}} \:=\:  -
\dfrac{N_c}{V^{(4)}} \sum_{f=u,d,s} \int d^4x \, d^4x' \
\trD \, \ln \left[ \mathcal{S}_f(x,x')\right]^{-1} -\dfrac{1}{2}
\left( \bar{\sigma}_f \ \bar{\mbox{s}}_f + G\ \bar{\mbox{s}}_f \
\bar{\mbox{s}}_f - \dfrac{K}{2}\, \bar{\mbox{s}}_u \
\bar{\mbox{s}}_d \ \bar{\mbox{s}}_s \right) \ ,
\label{seff}
\end{align}
where $\trD$ stands for the trace in Dirac space and $\left[\mathcal{S}_f(x,x')\right]^{-1} = \delta(x-x')\left[ -i ( \slashed \partial - i Q_f \slashed {\cal A} ) + M_f \right]$ is the inverse mean field quark propagator for each flavor, with effective mass $M_f=m_f+\bar{\sigma}_f$.
In the latter equation, $\bar{\mbox{s}}_f = \tilde{\mbox{s}}_f(\bar \sigma_a)$ represents the auxiliary fields at the mean-field level within the SPA approximation (note that $\bar{\mbox{p}}_f=0$).
From the condition $\delta \bar{S}_\sl[\rm E]^\sl[\,\rm bos] / \delta \bar{\sigma}_f=0$ it follows that 
$\bar{\mbox{s}}_f = 2 \phi_f$, where $\phi_f$ is the chiral condensate for each flavor given by
\begin{align}
\phi_f \:=\: \langle{\bar \psi}_{f} \psi_{f}\rangle \:=\: -\dfrac{\delta \bar{S}_\sl[\rm E]^\sl[\,\rm bos]}{\delta m_f} \:=\: 
- \dfrac{N_c}{V^{(4)}} \int d^4x \ \trD \ \mathcal{S}_f(x,x) \ .  
\label{cond_schwPT}
\end{align}
After expressing the quark propagator $\mathcal{S}_f(x,x')$ in its Schwinger proper-time form, see Eq.~\eqref{sfx}, the extension to finite temperature is performed through the replacements
\begin{align}
\int \dfrac{dp_4}{2\pi}  f(p_4) \ \rightarrow \ T \sum_{\ell=-\infty}^\infty f(\omega_\ell) \quad ; \quad 
p_4 \ \rightarrow \ \omega_\ell \:=\: (2 \ell +1)\pi T \ ,
\label{Trep}
\end{align}
in the magnetic quark condensate, where $\omega_\ell = (2 \ell+1) \pi T$ are the fermionic Matsubara frequencies~\cite{bailin1993introduction}.
In the spirit of the magnetic field independent regularization (MFIR) scheme, by adding and subtracting the $B=0$ contribution each regularized chiral condensate can be written as
\begin{align}
\phi_f^{B,T} \:=\: \phi_f^{0,T} + \phi_f^{mag,T} \qquad \rightarrow \qquad
\begin{cases}
\phi_f^{0,T}   \:\equiv\: -N_c M_f \, I_{1f}^{0,T} \\[1mm]
\phi_f^{mag,T} \:\equiv\: -N_c M_f \, I_{1f}^{mag,T}
\end{cases} \ ,
\label{phif-m}
\end{align}
where $\phi_f^{0,T}$ represents its non-magnetic contribution (although implicitly dependent on $B$ through $M_f$) and 
$\phi_f^{mag,T}$ is the corresponding thermomagnetic contribution. 
Definitions of the $I_{1f}$ functions as well as details of the condensate calculation are presented in Appendix~\ref{app-B}.
Nonetheless, we would like to remark a subtle point about the $B=0$ function $I_{1f}^{0,T}$.
By properly manipulating the thermal sum over Matsubara modes, this function can be further decomposed into vacuum and thermal integrals
\begin{align}
I_{1f}^{0,T} \:=\: I^{vac \vphantom{p}}_{1f} + I^{0,ther}_{1f} \ .
\end{align}
In principle, only the vacuum ($B=T=0$) contribution is divergent and thus required to be regularized.
In contrast, the thermal contribution 
\begin{align}
I^{0,ther}_{1f} \:=\: -\dfrac{2}{\pi^2} \int dp\: p^2 \: \dfrac{1}{E_f} \, \dfrac{1}{1+e^{E_f/T}} \ ,
\end{align}
with $E_f=\sqrt{p^2+M_f^2}$, is convergent, i.e. the integral can be performed from $p=0$ up to infinity.
The choice of whether regularize or not the thermal contribution depends on the physical context. 
When integrating up to $\Lambda$, the effective quark masses correctly tend to their current values at high temperatures~\cite{Costa:2007fy,Costa:2009ae,Costa:2010zw,Xue:2021ldz,Florkowski:1993br}, leading to vanishing quark condensates and the restoration of chiral symmetry. 
Nevertheless, the analysis in different versions of the NJL model indicate that the usual Stefan-Boltzmann limit is not reached for several thermodynamical quantities~\cite{Zhuang:1994dw,Costa:2007fy,Costa:2009ae,Avancini:2020xqe,Pasqualotto:2023hho,Xue:2021ldz}. 
In contrast, while integrating up to infinity leads to a good behavior for the latter quantities, it spoils the $T\rightarrow \infty$ behavior of $M_f$ which become lower than $m_f$, resulting in positive values for the condensates.
Therefore, in both methods it is difficult to avoid all subtleties when dealing with thermal integrations at high temperatures, and some attempts have been applied over the years to solve this issue~\cite{Moreira:2010bx,Bratovic:2012qs}.
Since in this work we study meson masses, which depend directly on effective quarks masses, we will choose to regularize thermal contributions. 
As stated in Refs.~\cite{Florkowski:1993br,Florkowski:1993bq}, this choice also leads to a good asymptotic behavior $m_\sl[scr]\rightarrow 2\pi T$ for screening masses at high temperatures.

Finally, inserting the constraint $\bar{\mbox{s}}_f = 2 \phi_f$ from the gap equation into the SPA conditions of Eq.~\eqref{SPA}, we obtain the following set of regularized coupled equations for the effective quark masses
\begin{align}
 M_{u} & \:=\: m_u - 4 G\ \phi_u^{B,T} + 2K\ \phi_d^{B,T}\, \phi_s^{B,T} \ , \nonumber\\[2mm]
 M_{d} & \:=\: m_d - 4 G\ \phi_d^{B,T} + 2K\ \phi_s^{B,T}\, \phi_u^{B,T} \ , \nonumber\\[2mm]
 M_{s} & \:=\: m_s - 4 G\ \phi_s^{B,T} + 2K\ \phi_u^{B,T}\, \phi_d^{B,T} \ .
\label{gapeqs}
\end{align}
Since $\phi_f$ depends on $M_f$, the effective quark masses have to be self-consistently calculated. 

\vspace*{2mm}
\subsection{Meson sector}

Mesons are described as field fluctuations around mean-field values. 
In particular, their masses arise from quadratic-order fluctuations.
Expanding the bosonized Euclidean action one gets 
\begin{align}
S^\sl[\rm bos]_\sl[\rm E] \:=\: \bar{S}^\sl[\rm bos]_\sl[\rm E] + S^{\sl[\rm quad]}_{\rm mes} \ ,
\end{align} 
where $\bar{S}^\sl[\rm bos]_\sl[\rm E]$ is the mean-field action given in Eq.~\eqref{seff}, while for the pseudoscalar sector
\begin{align}
S^\sl[\rm quad]_\sl[\rm mes] \:=\: \dfrac{1}{2} \, \sum_{PP'} \ \int d^4 x' d^4x \
\delta P(x)^\dagger \ {\cal G}_{PP'}(x,x')\ \delta P'(x') \ ,
\label{Gdiag}
\end{align}
where in the summation $P,P'=\pi_3, \pi^\pm, K^0, \bar K^0,$ $K^\pm, \eta_0, \eta_8$ are associated with the nonet of pseudoscalar mesons. 
In the latter expression, ${\cal G}_{PP'}(x,x')$ stands for the inverse meson propagator in coordinate space and can be written as
\begin{align}
{\cal G}_{PP'}(x,x') \:=\: T_{PP'} \ \delta^{(4)}(x-x') - J_{PP'}(x,x') \ .
\label{Gprop}
\end{align}
For $P,P'=\pi^\pm, K^\pm, K^0, \bar K^0$ this operator is diagonal
\begin{align}
T_{PP'} \:=\: T_\sl[P] \ \delta_{PP'} \qquad ; \qquad J_{PP'}(x,x') \:=\: J_\sl[P](x,x') \ \delta_{PP'} \ ,
\end{align}
where the polarization functions can be written as
\begin{alignat}{2}
J_{\pi^+}(x,x') &\:=\: J_{\pi^-}(x',x) &&\:=\:  c_{ud}(x,x') \ , \nonumber \\[2mm]  
J_{K^+}(x,x') &\:=\: J_{K^-}(x',x) &&\:=\: c_{us}(x,x') \ , \nonumber \\[2mm]
J_{K^0}(x,x') &\:=\: J_{\bar K^0}(x',x) &&\:=\: c_{ds}(x,x') \ .
\end{alignat}
In contrast, ${\cal G}_{PP'}(x,x')$ is non-diagonal but symmetric in the $P,P'=\pi_3, \eta_0, \eta_8$ subspace
\begin{align}
{\cal G}_{PP'}(x,x') \:=\: T_{PP'} \ \delta^{(4)}(x-x') - \sum_f
\gamma^f_{PP'} \ c_{ff}(x,x') \ ,
\label{Gprop_mix}
\end{align}
leading to $\pi^0-\eta-\eta'$ mixing.
The values of $T_{PP'}$ and $\gamma^f_{PP'}$ can be found in Appendix~\ref{app-A}.
In the previous expressions, we have defined the reduced polarization functions 
\begin{align}
c_{ff'}(x,x') \:=\: 2 N_c \ \trD \left[ \mathcal{S}_f(x,x') \ \gamma_5 \  \mathcal{S}_{f'}(x',x) \ \gamma_5 \right] \ .
\label{cff'}
\end{align}
The latter are the key ingredient for the calculation of meson screening masses.

In order to diagonalize the mesonic action of Eq.~\eqref{Gdiag}, we expand meson fields as
\begin{align}
\delta P(x) \:=\: \sumint_{\bar q} \: \mathcal{F}_Q (x,\bar q) \; \delta P(\bar q) \ ,
\label{transf}
\end{align}
where the functions $\mathcal{F}_Q (x,\bar q)$ are solutions of the meson field equations in the presence of an external constant and homogeneous magnetic field, with associated quantum numbers specified by $\bar q$.
As discussed in following subsections, the definition of the ``integral" symbol over $\bar{q}$ depends on whether the meson is neutral or charged. 
As a result, the transformed reduced polarization functions will be given by
\begin{align}
\mathcal{C}_{ff'}(\bar q,\bar q') \:=\:  
\int d^4x' d^4x \ \mathcal{F}_Q(x,\bar q)^\ast \: c_{ff'}(x,x') \: \mathcal{F}_Q(x',\bar q') \ .
\label{cff'_red}
\end{align}
Details of their computation are given in Appendix~\ref{app-D}.
Note that for the calculations it is convenient to consider first the $T=0$ case of the expressions, and then perform the extension to finite temperature through the replacements defined in Eq.~\eqref{Trep} together with $q_4 \rightarrow \nu_n = 2n\pi T$ for the meson $4-$momentum.
Since we are interested in the determination of screening masses $m_{\rm scr}$ associated with the meson lowest Matsubara mode $n=0$, we will set $\nu_n=0$ in what follows.


\vspace*{3mm}
\subsubsection{Neutral mesons}

For neutral mesons, Schwinger phases arising from equal flavor quark propagators in Eq.~\eqref{cff'} cancel out.
As a result, the polarization functions only depend on the coordinate difference $x-x'$, i.e. they are translationally invariant.
At zero temperature, they are diagonalized by a usual Fourier transformation $\mathcal{F}_Q(x,\bar q)=\exp(iq x)$, where $\bar q=q=(\vec q,q_4)$ stands for the usual four-momentum.
The quadratic meson action in the momentum basis will be then given by
\begin{align}
S^{\sl[\rm quad]}_{\sl[\rm neut.mes]} \:=\ & \dfrac{1}{2} \ \ \sum_{P=K^0,\bar K^0} \:\quad 
\int \dfrac{d^4q}{(2\pi)^4} \ \delta P^*(-q)\ {\cal G}_{P}(q_{\per}^2,q_{\npar}^2)\ \delta P(q) \ +
\nonumber \\[2mm]
& \dfrac{1}{2} \sum_{P,P'=\pi_3,\eta_0,\eta_8} \ 
\int \dfrac{d^4q}{(2\pi)^4} \ \delta P^\ast(-q)\ {\cal G}_{PP'}(q_{\per}^2,q_{\npar}^2)\ \delta P'(q) \ ,
\end{align}
where $q_{\per}=(q_1,q_2)$ and $q_{\npar}=(q_3,q_4)$.
Here, the inverse neutral kaon propagator is simply
\begin{align}
{\cal G}_{K^0}(q_{\per}^2,q_{\npar}^2) \:=\: {\cal G}_{\bar K^0}(q_{\per}^2,q_{\npar}^2) \:=\:
\left( 2 G - K \phi_u \right)^{-1} - c_{ds}(q_{\per}^2,q_{\npar}^2) \ , 
\label{GK0}
\end{align}
while for $P,P'=\pi_3,\eta_0,\eta_8$ we have
\begin{align}
{\cal G}_{PP'}(q_{\per}^2,q_{\npar}^2) \:=\: T_{PP'}- \sum_f \gamma^f_{PP'}\ c_{ff}(q_{\per}^2,q_{\npar}^2) \ . 
\label{Gff}
\end{align}
The values of $T_{PP'}$ and $\gamma^f_{PP'}$ can be found in Eqs.~\eqref{eq8} and~\eqref{gammas}, respectively.

Finally, extending to finite temperature via Eq.~\eqref{Trep}, we can determine the neutral meson screening masses from the inverse propagators at finite temperature and magnetic field ${\cal G}^{B,T}_{K^0}(q_{\per}^2,q_3^2)$ and ${\cal G}^{B,T}_{PP'}(q_{\per}^2,q_3^2)$.
The expression of $c_{ff'}^{B,T}(q_{\per}^2,q_{\npar}^2)$ regularized within the MFIR scheme can be found in Eq.~\eqref{cff'neutro}, where the thermomagnetic function is given by a trivial generalization of the $T=0$ expression found in Ref.~\cite{Avancini:2021pmi}.
Due to the anisotropy induced by the magnetic field, there are two screening masses for each neutral meson: the perpendicular $m_{\sl[scr,\,\per]}$ and the parallel screening mass $m_{\sl[scr,\npar]}$. 
For $K^0$ and $\bar K^0$, they are determined by the roots of
\begin{align}
{\cal G}^{B,T}_{K^0}(-m^2_{\sl[scr,\,\per]},0) \:=\:0 \quad ; \quad  
{\cal G}^{B,T}_{K^0}(0,-m^2_{\sl[scr,\npar]}) \:=\:0 \ .
\end{align}
Similarly, defining
\begin{align}
\mathcal{M}^{B,T} \:=\: \begin{pmatrix}
\mathcal{G}^{B,T}_{\pi_3\pi_3} & \mathcal{G}^{B,T}_{\pi_3\eta_0} & \mathcal{G}^{B,T}_{\pi_3\eta_8} \\
\mathcal{G}^{B,T}_{\eta_0\pi_3} & \mathcal{G}^{B,T}_{\eta_0\eta_0} & \mathcal{G}^{B,T}_{\eta_0\eta_8} \\
\mathcal{G}^{B,T}_{\eta_8\pi_3} & \mathcal{G}^{B,T}_{\eta_0\eta_8}
& \mathcal{G}^{B,T}_{\eta_8\eta_8} \end{pmatrix} \ ,
\label{Mprop}
\end{align}
the complex masses of the $\pi^0-\eta-\eta'$ mesons are associated with the roots of
\begin{align}
\det[\mathcal{M}^{B,T}(-m^2_{\sl[scr,\,\per]},0)] \:=\:0 \quad ; \quad
\det[\mathcal{M}^{B,T}(0,-m^2_{\sl[scr,\npar]})] \:=\: 0 \ .
\label{eqnneutral}
\end{align}

\vspace*{3mm}
\subsubsection{Charged mesons}

For charged mesons, the Schwinger phases appearing in the polarization functions do not cancel out, leading to a breakdown of translational invariance.
To overcome this difficulty, the Ritus formalism is used.
Thus, at $T=0$, the function $\mathcal{F}_Q(x,\bar q)$ of the charged meson field expansion~\eqref{transf} is a gauge dependent Ritus-like function with quantum numbers $\bar q=(k,\chi,q_3,q_4)$, where $k$ is a non-negative integer related to the so-called Landau level while $\chi$ can be chosen according to the gauge fixing~\cite{Dumm:2023owj,Wakamatsu:2022pqo}. 
A throughout discussion of the technical details is deferred to Appendix~\ref{app-C}, while the explicit form of $\mathcal{F}_Q(x,\bar q)$ for standard gauges can be found in Ref.~\cite{Dumm:2023owj}. 
In the Ritus basis, the charged meson contribution to the quadratic action can be written as 
\begin{align}
S^{\sl[\rm quad]}_{\sl[\rm char.mes]} \:=\: \dfrac{1}{2} \: \sum_{P=\pi^\pm, K^\pm} \:  \sumint_{\bar{q}} \: 
\delta P^*(\bar{q})\ {\cal G}_{P}(k,\Pi^2)\ \delta P(\bar{q}) \ ,
\end{align}
where $\Pi^2 = q_{\npar}^2 + (2k + 1)B_\sl[P]$ with $B_\sl[P]=|Q_\sl[P] B|$. 
In particular, the ``integral" symbol over $\bar{q}$ will depend on the chosen gauge. 
For definiteness, in the Landau gauge ${\cal A}_\mu=\delta_{\mu 2} x_1 B$ it is given by 
\begin{align}
\sumint_{\bar{q}} \:=\: \dfrac{1}{2\pi} \: \sum_{k=0}^\infty \: \int \, \dfrac{dq_2dq_3dq_4}{(2\pi)^3} \ .    
\end{align}
The inverse propagators read
\begin{align}
{\cal G}_{\pi^\pm}(k,\Pi^2) & \:=\: \left[ 2 G - K \phi_s \right]^{-1} - c_{ud}(k,\Pi^2) \ , \nonumber\\[2mm]
{\cal G}_{K^\pm}(k,\Pi^2) & \:=\: \left[ 2 G - K \phi_d \right]^{-1} - c_{us}(k,\Pi^2) \ .
\label{gcharge-m}
\end{align}
Once again, using Eq.~\eqref{Trep} to extend to finite temperature (at $q_4\rightarrow \nu_n=0$) we obtain the thermomagnetic inverse propagators ${\cal G}^{B,T}_{\pi^\pm}(k,\Pi^2)$ and ${\cal G}^{B,T}_{K^\pm}(k,\Pi^2)$.
The expression of $c_{ff'}^{B,T}(k,\Pi^2)$ regularized within the MFIR scheme can be found in Eq.~\eqref{cff'carg}.
Differently from the neutral case, for charged mesons only the parallel screening masses $m_{\sl[scr,\npar]}$ can be defined, for each Landau level. 
They are obtained as the roots of the equations
\begin{align}
{\cal G}^{B,T}_{\pi^\pm}(k,-m_{\sl[scr,\npar]}^2) \:=\:0 \quad ; \quad {\cal G}^{B,T}_{K^\pm}(k,-m_{\sl[scr,\npar]}^2) \:=\: 0 \ .
\end{align}


\vspace*{6mm}
\section{Numerical results}\label{Numerical_Results}

\subsection{Model parametrization and critical temperatures}


To obtain numerical results for the magnetic field dependence of meson masses one has to fix the model parametrization. 
Here, following Ref.~\cite{Rehberg:1995kh}, we take the parameter set 
$m_u = m_d = 5.5$~MeV, $m_s = 140.7$~MeV, $\Lambda = 602.3$~MeV, $G\Lambda^2 = 1.835$ and $K\Lambda^5=12.36$, which has been determined on fixing that for $B=T=0$ one gets $m_\pi=135$~MeV, $m_K=497.7$~MeV, $m_{\eta'}=957.8$~MeV and $f_\pi=92.4$~MeV.
This parameter set gives an $\eta$ mass of $m_\eta=514.8$~MeV, which compares reasonably well with the physical value $m^{\rm phys}_\eta= 548.8$~MeV, together with an appropriate value of the chiral condensate $\phi_f^{0,0}=(-242$~MeV$)^3$, for $f=u,d$.

As mentioned in the introduction, while local NJL-like models are able to reproduce the MC effect at vanishing temperature, they fail to lead to the IMC effect at $T \neq 0$. 
Among the possible ways to deal with this problem, one of the simplest consists of allowing the model parameters to depend on the magnetic field. 
Motivated by this approach, we explore the possibility of considering a magnetic field dependent coupling $G(B)$ \cite{Ferreira:2014kpa,Farias:2014eca,Farias:2016gmy} . 
We adopt the one proposed in Ref.~\cite{Ferreira:2014kpa} in the context of a similar SU(3) NJL model. 
In that work, the current quark masses, $\Lambda$ and $K$ were kept constant while for $G(B)$ the form 
\begin{align} 
G(B) \:=\: G \, \left[ \dfrac{ 1 + a \, (eB/\Lambda^2_\sl[\rm QCD])^2 + b \, (eB/\Lambda^2_\sl[\rm QCD])^3 }
                             { 1 + c \, (eB/\Lambda^2_\sl[\rm QCD])^2 + d \, (eB/\Lambda^2_\sl[\rm QCD])^4 } \right] \ , 
\label{fFCLFP}
\end{align}
was introduced, where $\Lambda_\sl[\rm QCD]=300$~MeV. 
As mentioned below, the parameters in $G(B)$ are determined by fitting the $B$-dependence of the pseudocritical temperature obtained in the model to the one found in the LQCD simulation of Ref.~\cite{Bali:2011qj}.

\begin{figure}[t!]
\centering{}\includegraphics[width=0.7\textwidth]{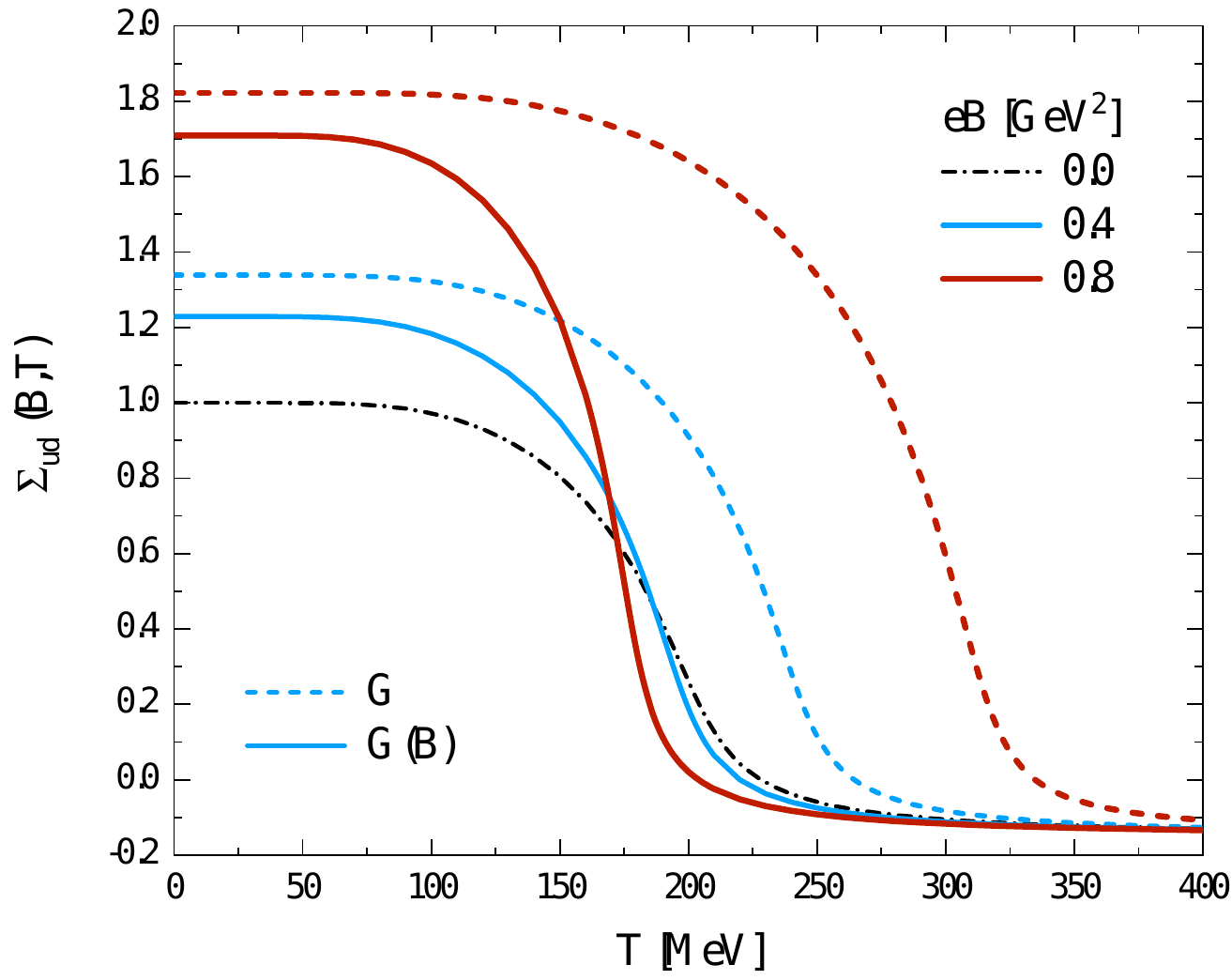}
\caption{The normalized average quark condensate $\Sigma_{ud}(B,T)$ as a function of temperature at fixed values of $eB$ (depicted by increasing thickness), for both constant (dashed lines) and magnetic field-dependent (solid lines) coupling constants.}
\label{fig:Sigud_fixB}
\end{figure}

\begin{figure}[t]
 \centering{}\includegraphics[width=0.9\textwidth]{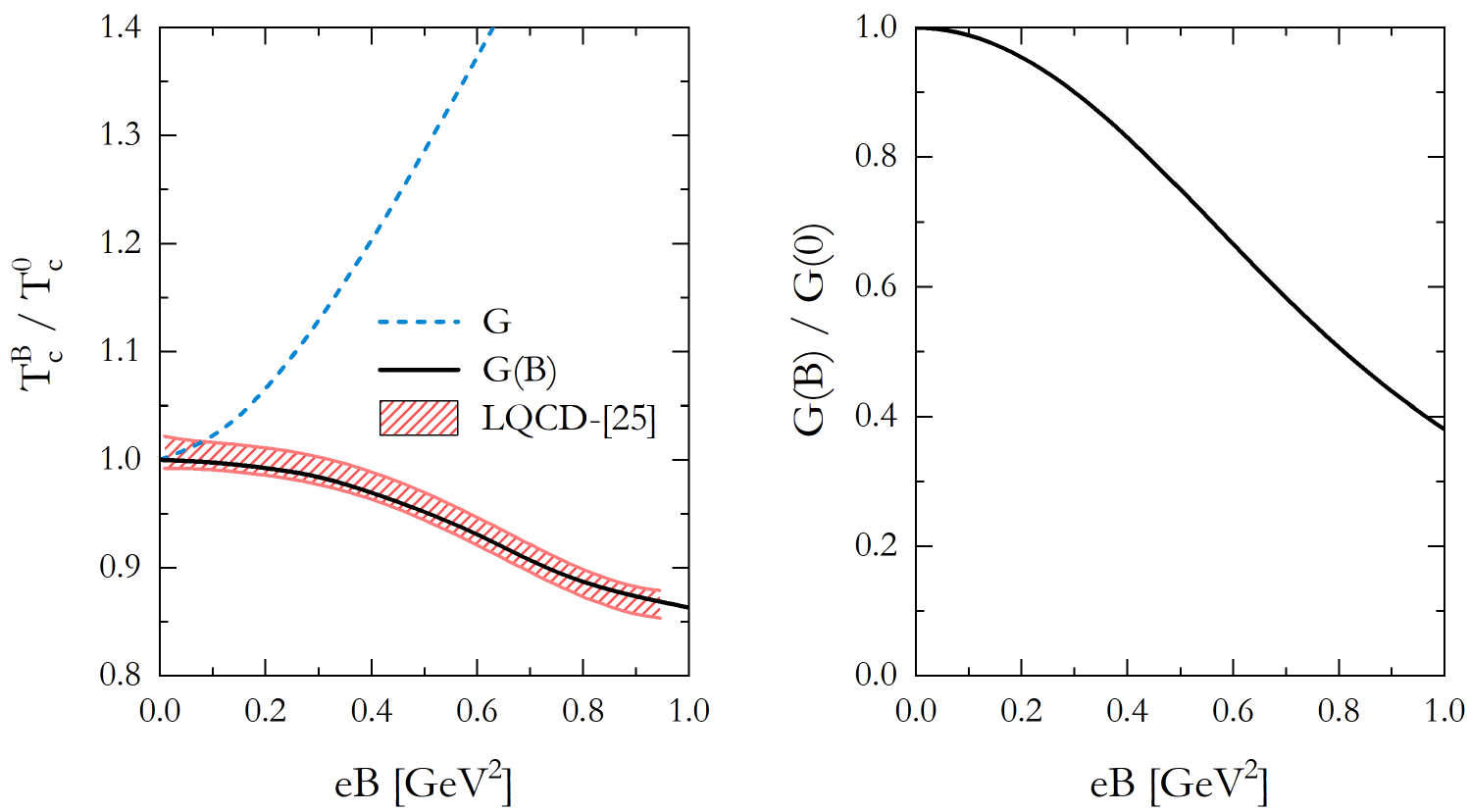}
\caption{Left panel: normalized pseudocritical temperature of the chiral transition as a function of $eB$ in our SU(3)-NJL model for both constant (dashed lines) and magnetic field-dependent (solid lines) coupling constants, together with LQCD results (red band)~\cite{Bali:2011qj}. 
LQCD data was normalized by $T_c^0|_\sl[\rm LQCD]=158$~MeV and NJL model results by $T_c^0=196.4$~MeV.
Right panel: normalized magnetic field-dependent coupling constant as a function of $eB$.}
\label{fig:Tc}
\end{figure}

The temperature dependence of the normalized average quark condensate $\Sigma_{ud}(B,T)$ for several values of the external field is shown in Fig.~\ref{fig:Sigud_fixB}. 
This quantity is defined as~\cite{Bali:2012zg} 
\begin{align}
\Sigma_{ud}(B,T) \:=\: \dfrac{m_u+m_d}{2D^2} \sum_{f=u,d} \left( \phi_f^{B,T}-\phi_f^{0,0} \right) + 1 \ ,
\end{align}
where $D =86$~MeV~$\times$~135~MeV. 
We consider first the result corresponding to $B=0$ (black dot-dashed line). 
From the inflection point of this curve we obtain that the chiral pseudocritical temperature at zero magnetic field is $T_c^0=196.4$~MeV, somewhat above the one found in LQCD,
$T_c^0|_\sl[\rm LQCD]=156$ MeV~\cite{HotQCD:2018pds}. 
To account for this difference, as often done in the literature, we will take $T_c^0$ as a reference value when discussing the temperature dependence of calculated quantities.
At finite $B$ we see that, when $T=0$, the expected increase of the condensate with the magnetic field (i.e., the Magnetic Catalysis effect) is realized for both constant $G$ (full lines) and $G(B)$ (dashed lines). 
On the other hand, while in the first case we observe an enhancement of the critical temperature with the magnetic field, for $G(B)$ the IMC effect is reproduced, as explicitly shown in the left panel of Fig.~\ref{fig:Tc}. 
The curves corresponding to the latter case have been obtained by fitting the LQCD pseudocritical temperature of Ref.~\cite{Bali:2011qj}, resulting in parameter values $a=0.0030882$, $b=1.716 \ 10^{-4}$, $c=0.0127782$ and $d=1.096 \ 10^{-4}$ for Eq.~\eqref{fFCLFP}.
The normalized magnetic coupling constant obtained is plotted in the right panel of Fig.~\ref{fig:Tc}.
It should be noticed that our values of $T_c^0$ and the parameters $a,b,c,d$ are somewhat different than those reported in Ref.~\cite{Ferreira:2014kpa}. 
This difference arises from the fact that, following the consistency arguments mentioned at the end of Sec.~\ref{sec2A}, we have introduced a 3D cutoff in the calculation of the thermal integral $I^{0,ther}_{1f}$ ---see Eq.~\eqref{i1temp}. 
For future reference we state that, for 0 GeV$^2\le eB \le 1$~GeV$^2$, the pseudocritical temperature lies in the range 196.4~MeV$\le T_c^B \le 342$~MeV for a constant $G$, while 169.5~MeV$\le T_c^B\le 196.4$~MeV for $G(B)$. 

\subsection{Neutral meson screening masses}

\begin{figure}[t!]
 \centering{}\includegraphics[width=0.8\textwidth]{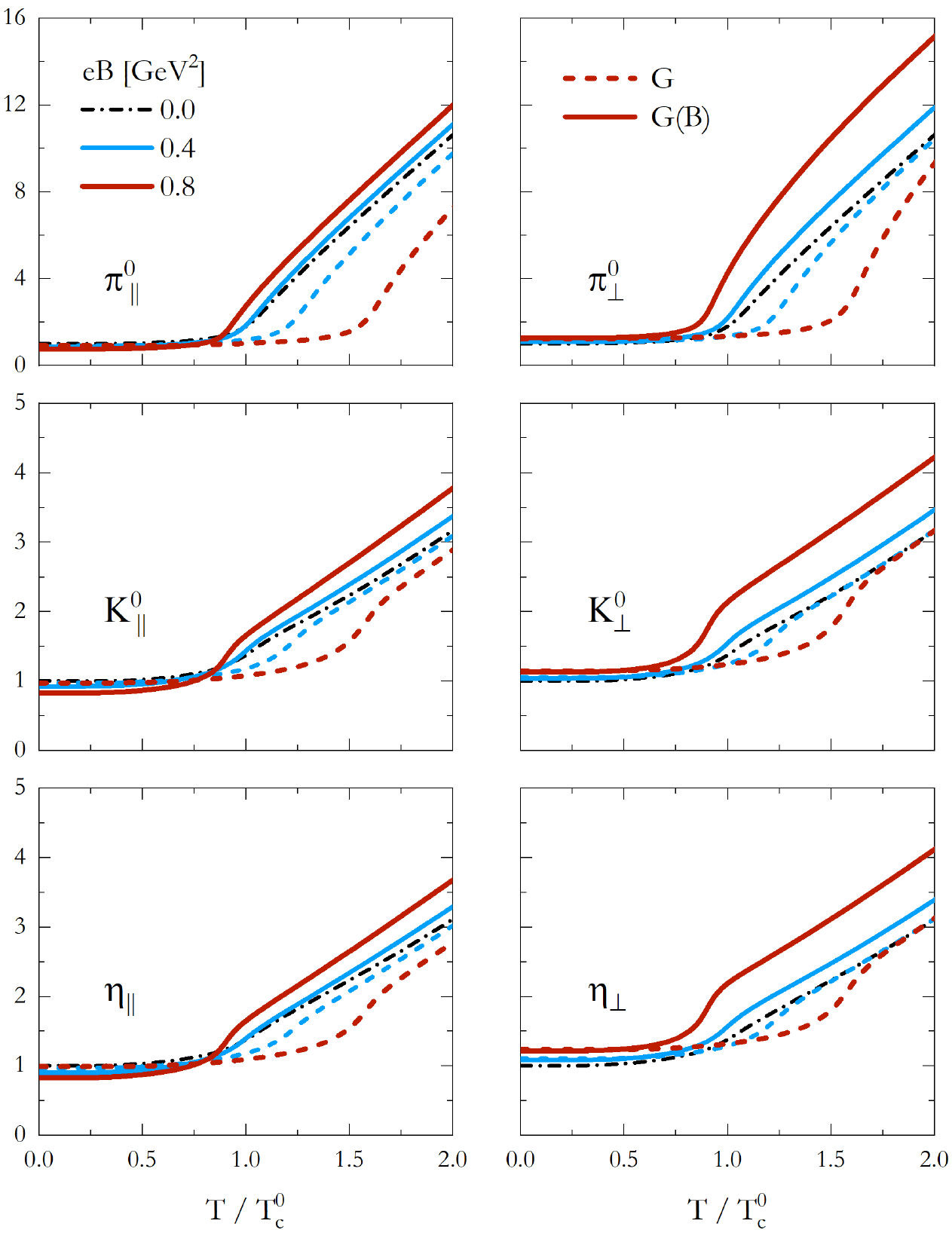}
\caption{Normalized screening masses $m_\sl[P](B,T)/m_\sl[P](0,0)$ of $P=\pi^0$ (top), $P=K^0$ (center) and $P=\eta$ (bottom) as
functions of the normalized pseudocritical temperature at fixed values of $eB$ (depicted by increasing thickness), 
for both constant (dashed lines) and magnetic field-dependent (solid lines) coupling constants. 
Left (right) panels correspond to parallel (perpendicular) screening masses.} 
\label{fig:m0NM_fixB}
\end{figure}

In this subsection we analyze the thermal (fix $B$) and magnetic (fix $T$) behavior of perpendicular and longitudinal screening masses of the neutral $\pi^0$, $K^0$ and $\eta$ mesons.
Regarding $\eta'$, at low temperatures its  mass lies above the $\bar{q}q$ decay threshold, characteristic of NJL-type models due to their lack of confinement, and thus develops an imaginary part related to its decay width.
At $T=0$, a careful study of the magnetic $\eta'$ complex parallel screening mass was performed in Ref.~\cite{Avancini:2021pmi}.
Although at $T\neq 0$ the magnetic $\eta'$ screening mass and width can be derived from Eq.~\eqref{eqnneutral} through analytic continuation, its analysis lies beyond the scope of our work.

In Fig.~\ref{fig:m0NM_fixB} we show results for the thermal dependence of the normalized screening masses of neutral mesons, at fixed values of $eB$. 
We see that all screening masses behave in a qualitatively similar way as functions of $T$. 
At $T<T_c^B$, they remain almost independent of the temperature. 
Then, around $T\sim T_c^B$, they undergo a transition rapidly increasing their values, which continue to increase (albeit more slowly) as $T$ grows. 
For a constant $G$ the transition occurs at larger temperatures for higher values of $eB$, as expected from the increase of $T_c^B$ seen in Fig.~\ref{fig:Tc}. 
The opposite is true for $G(B)$, reproducing the behavior seen in LQCD simulations~\cite{Ding:2022tqn}. 
Due to symmetry considerations and in order to satisfy the law of causality, we expect that $u_{\per}<u_{\npar}<1$ in a thermomagnetic medium, where $u_i$ is the meson sound velocity in the $i$-direction, and therefore $m_\sl[\rm scr,\npar] < m_\sl[\rm scr,\perp]$~\cite{Sheng:2020hge}.
This is checked in Fig.~\ref{fig:m0NM_fixB} for our results, where it is seen that perpendicular screening masses have higher values than longitudinal ones, an effect that is more pronounced for the pion case.

\begin{figure}[t!]
  \centering{}\includegraphics[width=0.8\textwidth]{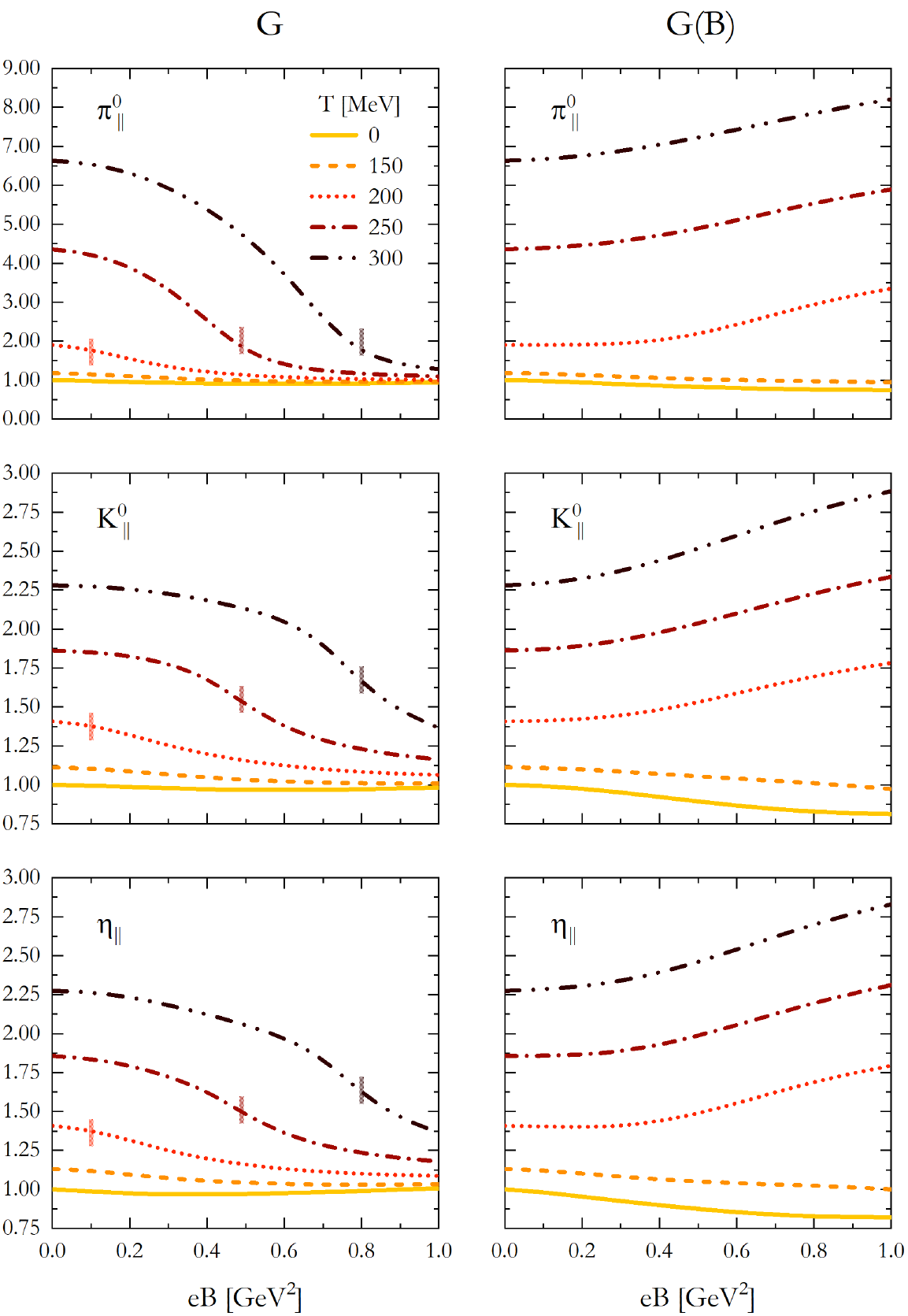}
\caption{Normalized longitudinal screening masses $m_\sl[P](B,T)/m_\sl[P](0,0)$ of $P=\pi^0_{\parallel}$ (top), $P=K^0_{\parallel}$ (middle) and $P=\eta_{\parallel}$ (bottom) as functions of $eB$ at fixed values of $T$, for both constant (left panels) and magnetic field-dependent (right panels) coupling constants.
Vertical lines indicate chiral transitions.}
\label{fig:m0_par_fixT}
\end{figure}

\begin{figure}[t!]
  \centering{}\includegraphics[width=0.8\textwidth]{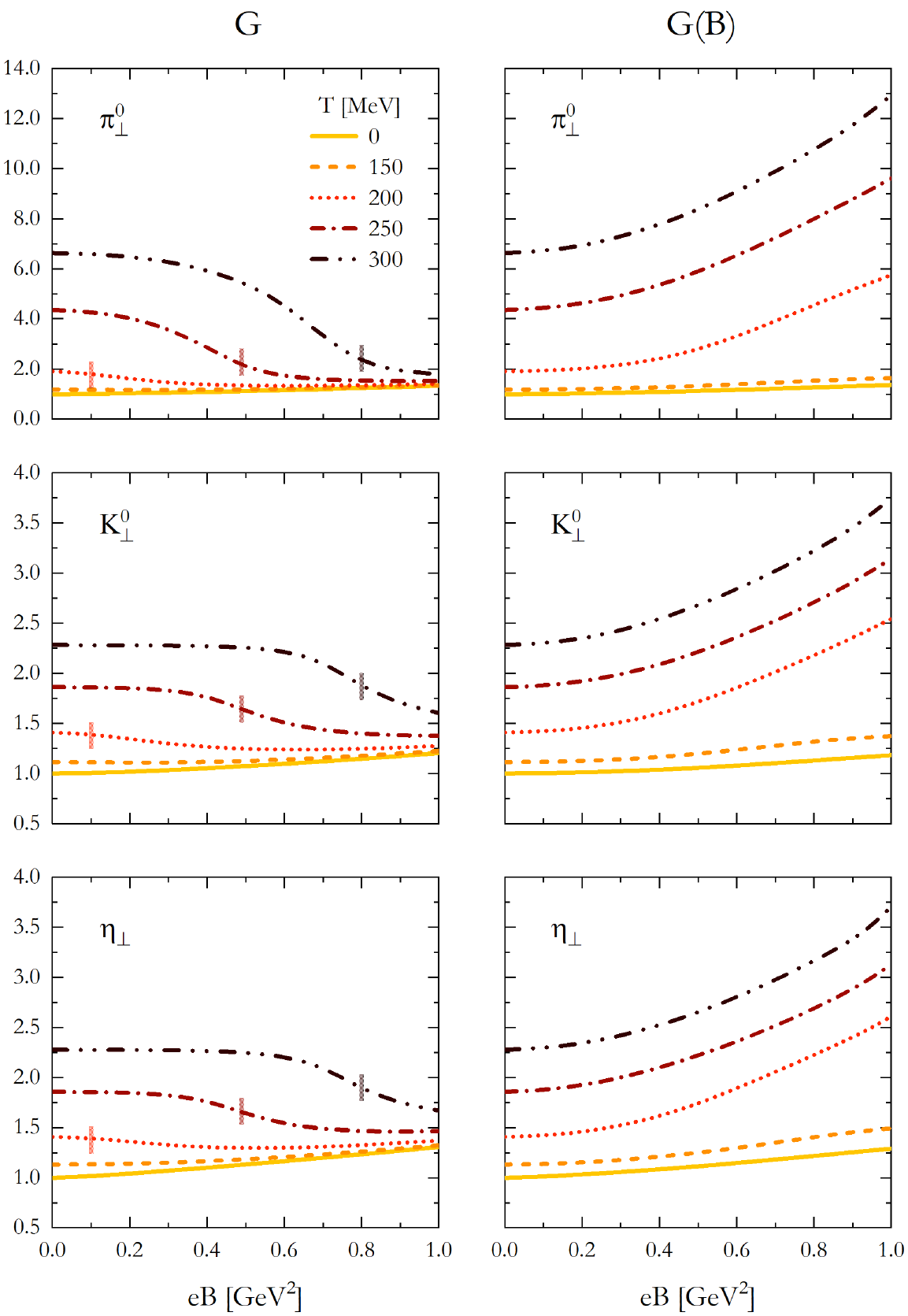}
\caption{Normalized perpendicular screening masses $m_\sl[P](B,T)/m_\sl[P](0,0)$ of $P=\pi^0_{\per}$ (top), $P=K^0_{\per}$ (middle) and $P=\eta_{\per}$ (bottom) as functions of $eB$ at fixed values of $T$, for both constant (left panels) and magnetic field-dependent (right panels) coupling constants.
Vertical lines indicate chiral transitions.}
\label{fig:m0_per_fixT}
\end{figure}

Similar plots but now at fixed values of $T$ are displayed in Figs.~\ref{fig:m0_par_fixT} and~\ref{fig:m0_per_fixT} for the $B$-dependence of longitudinal and perpendicular normalized screening masses, respectively.
Let us discuss the longitudinal case of Fig.~\ref{fig:m0_par_fixT} first.
At $T=0$, we obtain a non-monotonic behavior when using a constant coupling $G$ (not clearly visible in the graph due to the scale), while for $G(B)$ the curve becomes monotonically decreasing in good agreement with LQCD results~\cite{Ding:2021}, as discussed in Ref.~\cite{Avancini:2021pmi}.
At $T\neq 0$, the crossover points where $T=T_c^B$ are indicated by small vertical lines in the plot.
For $G(B)$, no vertical lines are displayed since the chosen values of temperature are always below or above $T_c^B$: 150~MeV$<T_c^B<200$~MeV.
For a constant $G$ and $T\gtrsim 150$~MeV all longitudinal masses decrease with the magnetic field for $eB<1$~GeV$^2$, undergoing a significant suppression in the transition region where $T\sim T_c^B$.
In contrast, for $G(B)$ their initial decreasing behavior reverses after $T_c^B$, where the masses are enhanced by $eB$.
These opposing magnetic behaviors originate in the also contrasting behavior of $T_c^B$, and can be better understood from Fig.~\ref{fig:m0NM_fixB}; at higher magnetic fields, $m_\sl[\mathrm{scr},\npar]$ starts to increase sooner (later) for $G(B)$ ($G$), thus reaching higher (lower) values at large temperatures.
On the other hand, LQCD simulations~\cite{Ding:2022tqn} show an intricate behavior; while around $T\sim T_c^B$ they also find an enhancement at large values of $eB$, as $T$ grows beyond $T_c^B$ this enhancement is suppressed, returning to a decreasing behavior at large values of $T$.
The disparity has two sides.
On one hand, there are uncertainties in the thermomagnetic dependence of $G$ in the local NJL model.
The LQCD behavior seems to lie somewhere between our choices of $G$ and $G(B)$.
Thus, choosing a softer functional dependence for $G(B)$ in our model would in principle lead to more similar results, although this inevitably leads to deviations on $T_c^B$, which was fitted to reproduce LQCD results.
On the other hand, in the lattice setup of Ref.~\cite{Ding:2022tqn} the vacuum pion mass $m_\pi=220$~MeV is somewhat higher than its physical value, which leads to a smaller IMC effect~\cite{DElia:2018xwo,Endrodi:2019zrl}.
In the NJL model, this would correspond to a softer functional dependence in $G(B)$.
If it is soft-enough, the decreasing trend found for a constant $G$ would be recovered.
Thus, one can expect lattice simulations at the physical point in the continuum limit to lead to a behavior of increased magnetic catalysis for the neutral pion longitudinal mass at large temperatures.
Nevertheless, \textit{a priori} it is not clear if the effects at the physical point will be enough to reverse the IMC-type behavior of $m_{\pi^0_{\npar}}$ seen in Ref.~\cite{Ding:2022tqn}, so as to provide the MC-type effect predicted by our particular choice of $G(B)$.
Despite the uncertainties in the $B,T$ dependence of $G$ in the local NJL model, our outcomes show that the use of $G(B)$ can help to obtain more consistent results with LQCD simulations, in contrast to constant-$G$ predictions which always lead to a magnetic decreasing behavior for $m_{\pi^0_{\npar}}$.

Regarding the perpendicular screening masses shown in Fig.~\ref{fig:m0_per_fixT}, at $T=0$ we find a magnetic enhancement of the masses, in contrast with the decreasing and non-monotonic behaviors found in the longitudinal case for $G(B)$ and $G$, respectively.
At finite temperatures, in the exhibited range 150~MeV~$\le T \le 300$~MeV, we observe a magnetic behavior similar to the one displayed by parallel masses: they increase (decrease) with $eB$ for $G(B)$ ($G$), again due to the behavior of $T_c^B$.
For higher temperatures, approximately $T \gtrsim 500$~MeV, the decreasing behavior found for $G$ reverses, becoming magnetically enhanced. 
This can already be seen from Fig.~\ref{fig:m0NM_fixB} for $K_{\per}^0$ and $\eta_{\per}$ mesons, where the $eB=0.4$~GeV$^2$ curve surpasses the $0.8$~GeV$^2$ one at the highest temperature shown.
This point will be made more clear in a following section regarding the high temperature limit of these masses.
One other subtlety should be remarked about Fig.~\ref{fig:m0_per_fixT}, related to the MFIR scheme. 
For perpendicular screening masses of neutral mesons, the full thermomagnetic $c_{ff'}^{B,T}$ functions of Eq.~\eqref{cff'neutro} are real for all values of $m_\sl[\rm scr,{\per}]$ (even at $T=0$), in contrast to the longitudinal case.
Nevertheless, within the MFIR one adds and subtracts the $B=0$ contribution~\eqref{cff'reg}, which develops an imaginary contribution when 
$m_\sl[\rm scr,{\per}] < |M_{f\ell} - M_{f'\ell}|$ or $m_\sl[\rm scr,{\per}] > M_{f\ell} + M_{f'\ell}$, where $M_{f\ell}=(M_f^2+\omega_\ell^2)^{1/2}$.
But these thresholds are artificial: the imaginary parts of the added and subtracted $B=0$ terms cancel out.
For the data displayed in Fig.~\ref{fig:m0_per_fixT}, this is relevant for $m_{\eta_{\per}}$, which overcomes the lower artificial threshold $2 (M_d^2+\pi^2 T^2)^{1/2}$ at $B\simeq 0.8~$GeV$^2$ when using $G(B)$.
The calculation of the remaining real contributions was performed through a generalization of the ideas presented in Ref.~\cite{Carlomagno:2022arc}.
We can also check from these plots that all values of perpendicular neutral meson masses are higher than those of longitudinal ones, as expected from causality. 
\linebreak


\subsection{Charged mesons}

In this subsection we discuss the thermomagnetic behavior of charged meson energies. 
Recall that, for the lowest Matsubara mode $n=0$, the charged parallel screening masses are defined by $\Pi^2=q_3^2+(2k+1)eB=-m_\sl[P]^2$, where $P=\pi^\pm_{\npar},\, K^\pm_{\npar}$.  
Thus, at the lowest Landau level $k=0$, one can define a charged parallel screening energy as $E_\sl[P]=\sqrt{m_\sl[P]^2+eB}$.

\begin{figure}[t!]
    \centering{}\includegraphics[width=0.9\textwidth]{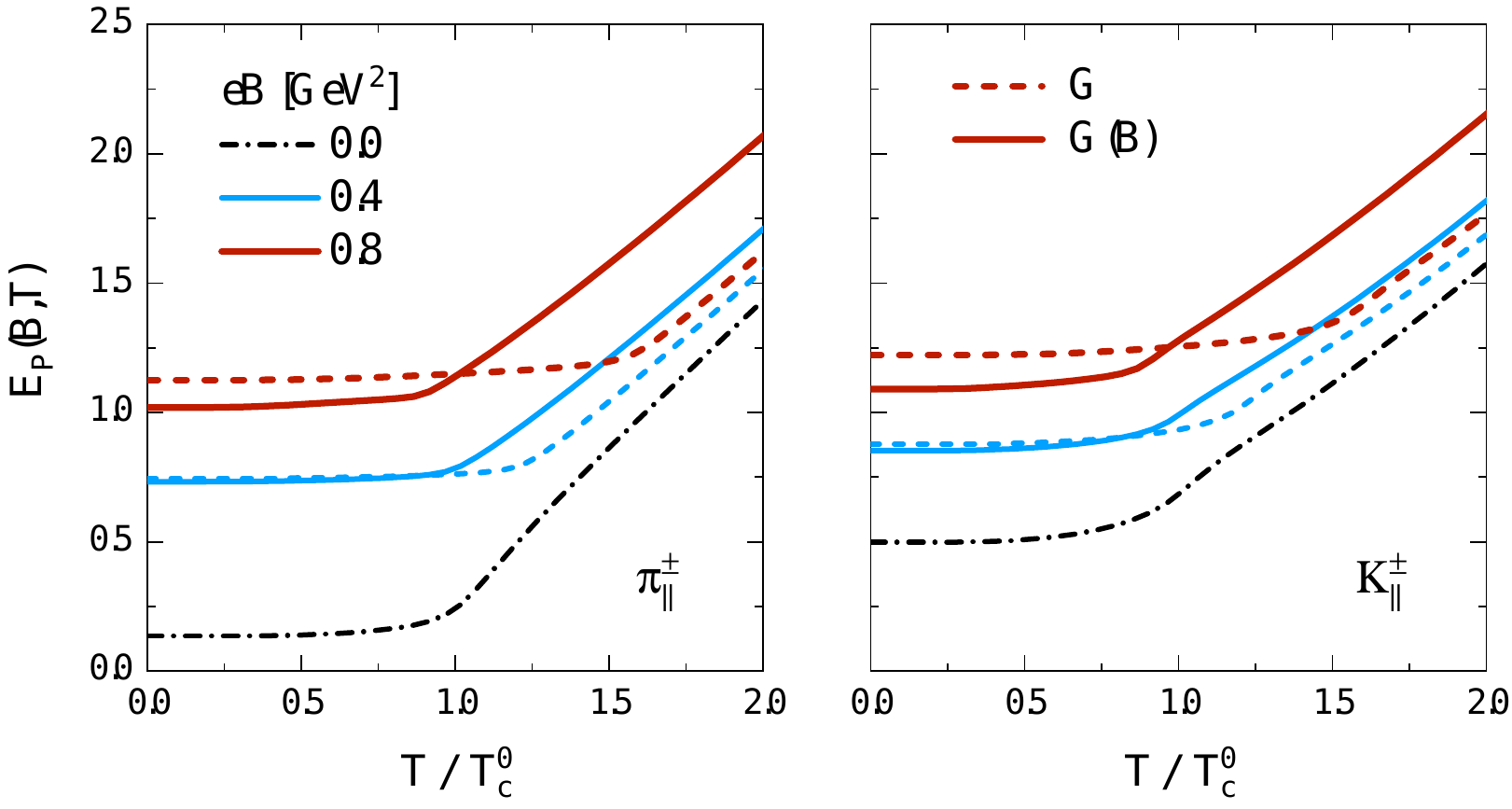}
\caption{Screening energies of charged pions (left) and kaons (right) as functions of the normalized pseudocritical temperature at fixed values of $eB$ (depicted by increasing thickness), for both constant (dashed lines) and magnetic field-dependent (solid lines) coupling constants.}
\label{fig:E_fixB}
\end{figure}

The temperature dependence of these charged screening energies is displayed in Fig.~\ref{fig:E_fixB} for fixed values of $eB$.
As seen, their thermal behavior is qualitatively similar to the neutral mesons one in Fig.~\ref{fig:m0NM_fixB}: while for $T<T_c^B$ the energies remain almost independent of the temperature, beyond $T_c^B$ they increase with $T$.
Nevertheless, the magnitude of the thermal enhancement seen at $T>T_c^B$ is moderately suppressed by the presence of the magnetic field.
Note that the vertical scale is the same for both mesons; at high magnetic fields the magnetic strength sets the scale and dominates over the meson mass scale.

\begin{figure}[t!]
    \centering{}\includegraphics[width=0.9\textwidth]{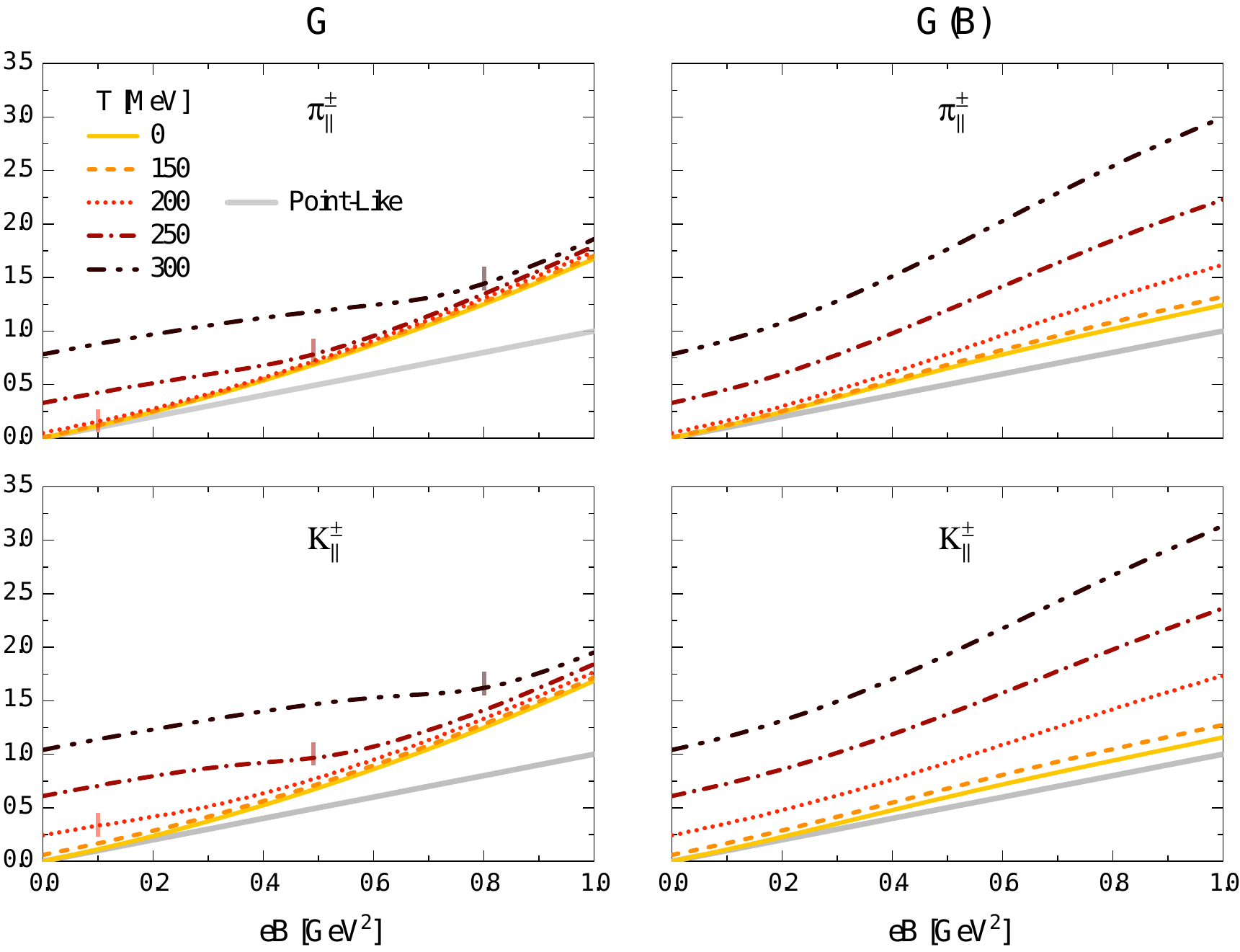}
\caption{Charged meson differences of squared lowest energies $E_\sl[P]^2(B,T)-E_\sl[P]^2(0,0)$ for charged pions (top) and kaons (bottom) as functions of $eB$ at fixed values of $T$, for both constant (left panel) and magnetic field-dependent (right panel) coupling constants.
The point-like case is depicted by a gray thick line.
Vertical lines indicate chiral transitions.}
\label{fig:E2_fixT}
\end{figure}

In Fig.~\ref{fig:E2_fixT} we focus on the magnetic behavior of the squared difference of charged screening energies, at fixed values of temperature. 
As expected from the explicit $eB$ dependence in the definition of the energy, we see an overall increase with the magnetic field for all temperatures.
For $T\lesssim 200$~MeV the use of a $B$-dependent coupling leads to lower values of energies as compared to using a constant $G$, thus lying more closely to the curve of the point-like case, although always above it.
This leads to a better agreement with $T=0$ results from LQCD simulations~\cite{Ding:2021} and NJL calculations with vector-axial mixing~\cite{Coppola:2023mmq}.
For $T\gtrsim 200$~MeV, the magnetic enhancement is more pronounced for $G(B)$.
This is closely related to the behavior of $T_c^B$.
For a constant $G$, $T_c^B$ grows with $B$, so when we increase $eB$ at fixed $T$ we can go from being above to being below the critical temperature; these transitions are indicated by vertical lines in the plot. 
Since we have seen in Fig.~\ref{fig:E_fixB} that below $T_c^B$ the energies are practically independent of $T$, the energy curves at high $T$ have to reduce the magnitude of their magnetic catalysis enhancement so as to approach the $T=0$ value.
In contrast, since $T_c^B$ diminishes with $B$ for $G(B)$, when $T\gtrsim 200$~MeV mesons always lie above the critical temperature in this case, and their energies can differ greatly from the $T=0$ values.

\subsection{High temperature limit}

\begin{figure}[t]
    \centering{}\includegraphics[width=0.6\textwidth]{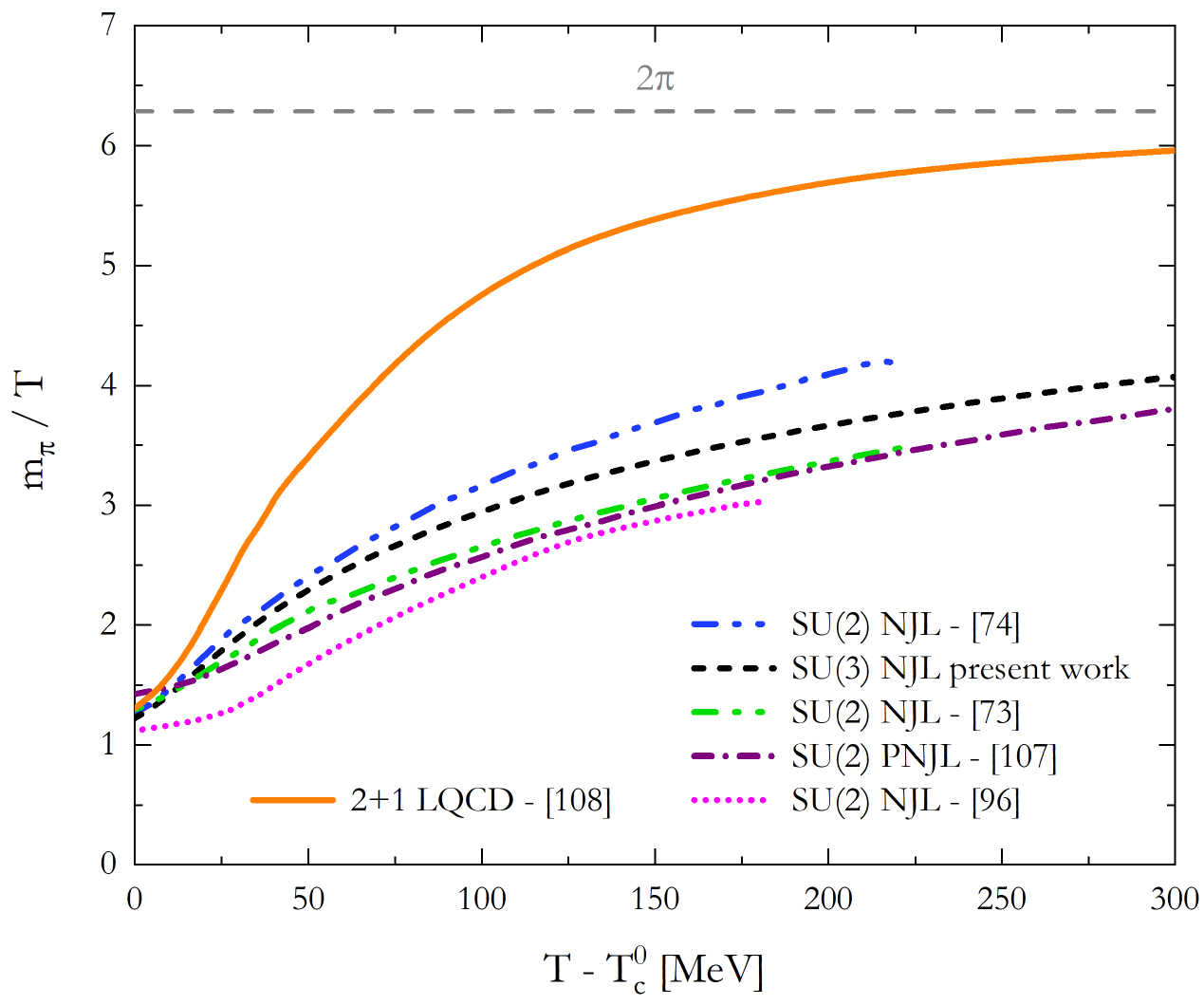}
\caption{Screening pion mass at $eB=0$ normalized by temperature as a function of the difference between temperature and the pseudocritical temperature at $eB=0$ for various versions of the local NJL models, including this work~\cite{Florkowski:1993br,Ishii:2013kaa,Sheng:2020hge,Sheng:2021evj}.
For comparison, the LQCD result of Ref.~\cite{Bazavov:2019www} is shown by a solid line.}
\label{fig:mpi_B0}
\end{figure}

In absence of external magnetic fields, it has been predicted that in the high temperature limit the meson screening masses should increase as 
$m_\sl[\rm scr] \sim 2\pi T$~\cite{Eletsky:1988an,Florkowski:1993bq}, as expected from the behavior of a gas of non-interacting massless quarks. 
Before considering the situation at finite magnetic field, it is interesting to analyze how this limit is approached within NJL-like models at $B=0$. 
In Fig.~\ref{fig:mpi_B0} we show predictions of several versions of the local NJL model for the pion screening mass normalized by the temperature as functions of the difference $T-T_c^0$. 
Here, $T_c$ is the critical temperature obtained in each particular calculation.
Given the rather broad range of values obtained for this quantity, we have chosen to subtract it so as to set a common reference scale. 
The versions of NJL models considered include our three-flavor model but also two-flavor models without~\cite{Florkowski:1993br,Sheng:2020hge,Sheng:2021evj} or with~\cite{Ishii:2013kaa} interactions with the Polyakov loop. 
It should be noticed that, in addition, different regularization schemes have been used among these works.
Also shown in Fig.~\ref{fig:mpi_B0} are predictions from the LQCD simulation of Ref.~\cite{Bazavov:2019www}.
Clearly, all versions of the NJL model lead to pion masses whose increase with temperature is somewhat slower than the one observed in LQCD results.
This means that the model overestimates the remaining interaction between quarks.


In spite of this overall overbinding, we believe that the analysis of the high temperature behavior of screening masses within the NJL model can provide valuable information on how such interactions are affected by the presence of an external magnetic field.  
First, we elaborate on the fact that its presence does not modify the expected asymptotic limit of $m_\sl[\rm scr] \sim 2\pi T$ at high temperatures. 
The energy of two non-interacting quarks in a thermomagnetic medium is given by 
\begin{align}
E_{2q} \:=\: \sqrt{[(2\ell_1+1)\pi T]^2+m_1^2+2\kappa_1|q_1 B|} + \sqrt{[(2\ell_2+1)\pi T]^2+m_2^2+2\kappa_2|q_2 B|} \ , 
\end{align}
where $\ell$ and $\kappa$ correspond to their Matsubara modes and Landau levels, respectively.
Thus, the lowest energy state (all others will be exponentially suppressed) is given by $\ell_{1,2}=\kappa_{1,2}=0$
\begin{align}
E_{2q}\bigr\rvert_\sl[\rm LES] \:=\: \pi T\, \sqrt{1+\left( \dfrac{m_1}{\pi T} \right)^2} +
                                     \pi T\, \sqrt{1+\left( \dfrac{m_2}{\pi T} \right)^2} 
                                     \quad  \xrightarrow[]{\ T\rightarrow \infty\ } \quad 2\pi T  \ .
\label{E2q_LES}
\end{align}
We see that there is no trace left of the magnetic field when taking the lowest Landau level.
As a consequence, all mesons share the same asymptotic limit, regardless of their charge or mass type (perpendicular or longitudinal). 
Moreover, corrections given by the $m_i/(\pi T)$ factor are negligible for the light sector under consideration; for the heaviest strange quark mass $m_s=95$~MeV, the correction amounts to only 1\% at $T=300$~MeV.

\begin{figure}[t!]
    \centering{}\includegraphics[width=0.8\textwidth]{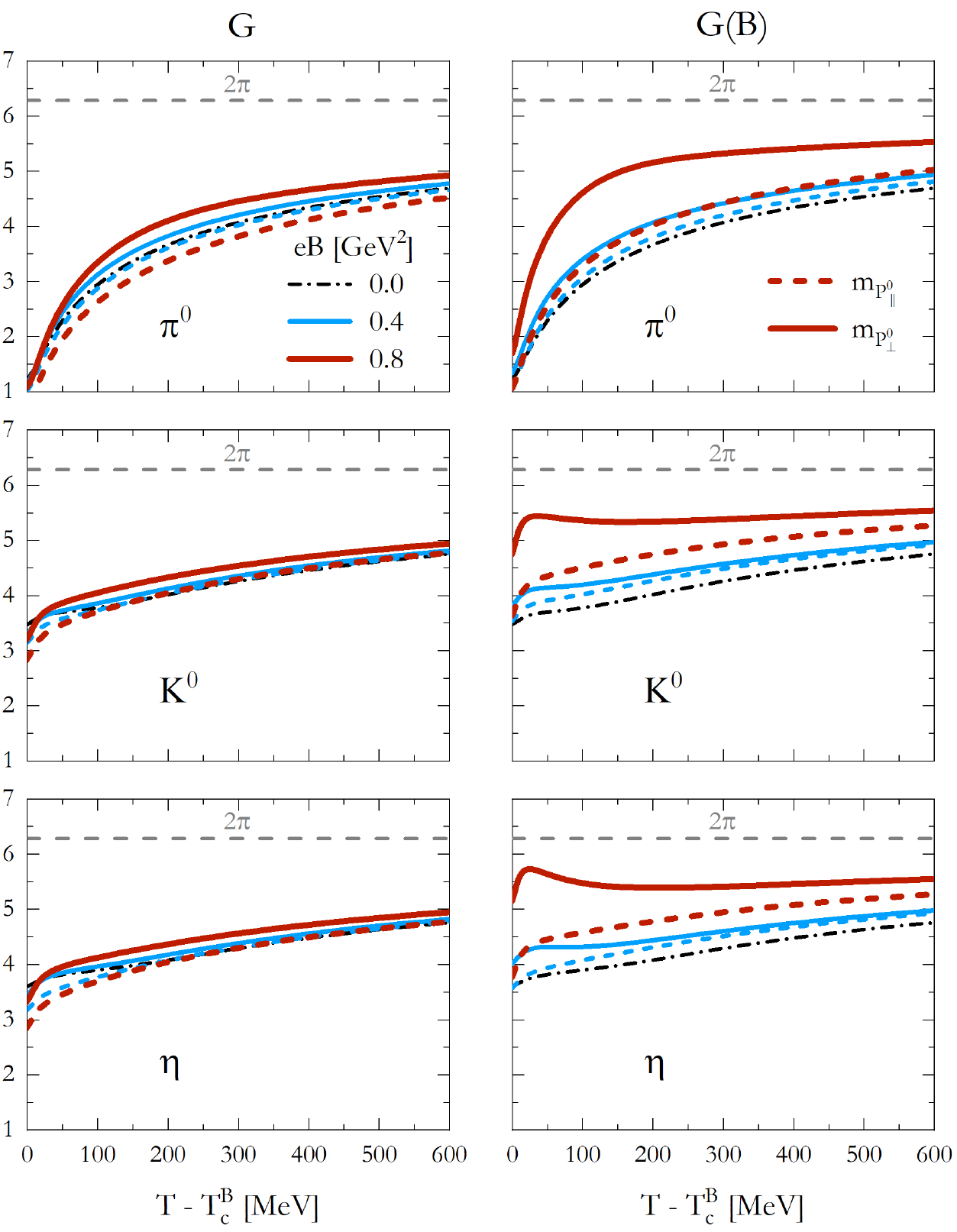}
\caption{Normalized longitudinal (dashed) and perpendicular (solid) screening masses $m_\sl[P](B,T)/T$ of $P=\pi^0$ (top), $P=K^0$ (center) and $P=\eta$ (bottom) as functions of the difference between temperature and the pseudocritical temperature, at fixed values of $eB$ (depicted by increasing thickness) for both constant (left) and $B$-dependent (right) coupling constants.}
\label{fig:m0-T_B}
\end{figure}

For neutral mesons, the high temperature dependence of their screening masses is shown in Fig.~\ref{fig:m0-T_B}.
The relation $m_\sl[\rm scr,\npar] < m_\sl[\rm scr,\perp]$ demanded by causality is still fulfilled~\cite{Sheng:2020hge}.
Regarding longitudinal masses, we see a contrasting behavior between the different couplings considered.
At a given value of $T-T_c^B$, the binding between quarks increases with the magnetic field for a constant $G$, while it decreases for $G(B)$, as in Ref.~\cite{Sheng:2021evj}. 
In a way, the latter behavior is to be expected since for $G(B)$ the coupling (and thus the interaction strength) decreases with $B$.
For perpendicular masses at high temperatures, the binding always decreases with $B$ for all couplings, in agreement with Ref.~\cite{Sheng:2021evj}.
As inferred from Figs.~\ref{fig:m0NM_fixB} and~\ref{fig:m0-T_B}, the magnetic field accelerates the thermal enhancement of perpendicular masses, leading to a magnetic catalysis at high temperatures.
For $G(B)$, this is seen from the higher mass values reached at the same temperature.
For $G$, even though at $eB=0.8$~GeV$^2$ perpendicular masses start increasing later compared to the $eB=0.4$~GeV$^2$ ones due to the displacement of $T_c^B$, see Fig.~\ref{fig:m0NM_fixB}, they overcome the $eB=0.4$~GeV$^2$ values when $T\gtrsim 500$~MeV, see Fig.~\ref{fig:m0-T_B}.
This magnetically catalyzed behavior at high-$T$ is not seen in Fig.~\ref{fig:m0_per_fixT} because the temperature in the display curves is below 500~MeV.

\begin{figure}[t]
    \centering{}\includegraphics[width=0.7\textwidth]{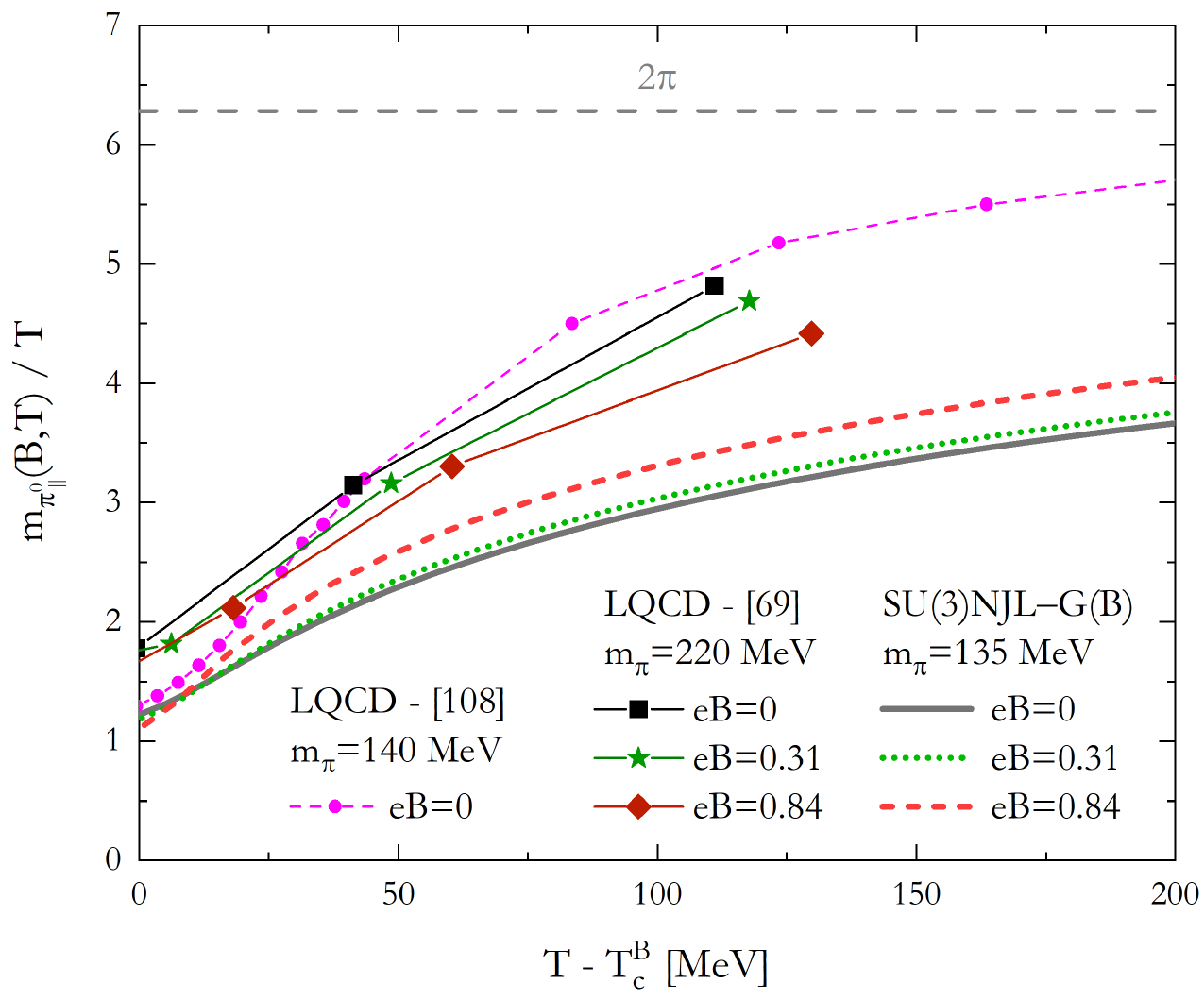}
\caption{Neutral pion longitudinal screening mass normalized by temperature as a function of the difference between temperature and the pseudocritical temperature, at fixed values of $eB$.
LQCD results of Refs.~\cite{Ding:2022tqn,Bazavov:2019www} are added for comparison, using their corresponding values of $T_c^B$.}
\label{fig:Comp-Ding}
\end{figure}

In Fig.~\ref{fig:Comp-Ding} we compare our results for the high temperature behavior of the $\pi^0_{\npar}$ screening mass  using $G(B)$ with the LQCD outcomes of Ref.~\cite{Ding:2022tqn}, at the same fixed values of $eB$. 
As seen, the curves behave in opposite ways as functions of the magnetic field: while in our case the binding decreases with $eB$, as in Ref.~\cite{Sheng:2021evj}, it increases for LQCD simulations. 
Similar trends are obtained for $K^0_{\npar}$.
This is related to the discussion around Fig.~\ref{fig:m0_par_fixT}, where it is stated that uncertainties arise from both approaches. 
If LQCD simulations at the physical point in the continuum limit find the same trend of increasing binding, this would imply that it is difficult to obtain both the IMC effect of $T_c^B$ and the correct high temperature behavior of $m_{\pi^0_{\npar}}$ within the NJL model by simply introducing a $B$-dependent coupling constant.

\begin{figure}[t]
    \centering{}\includegraphics[width=0.9\textwidth]{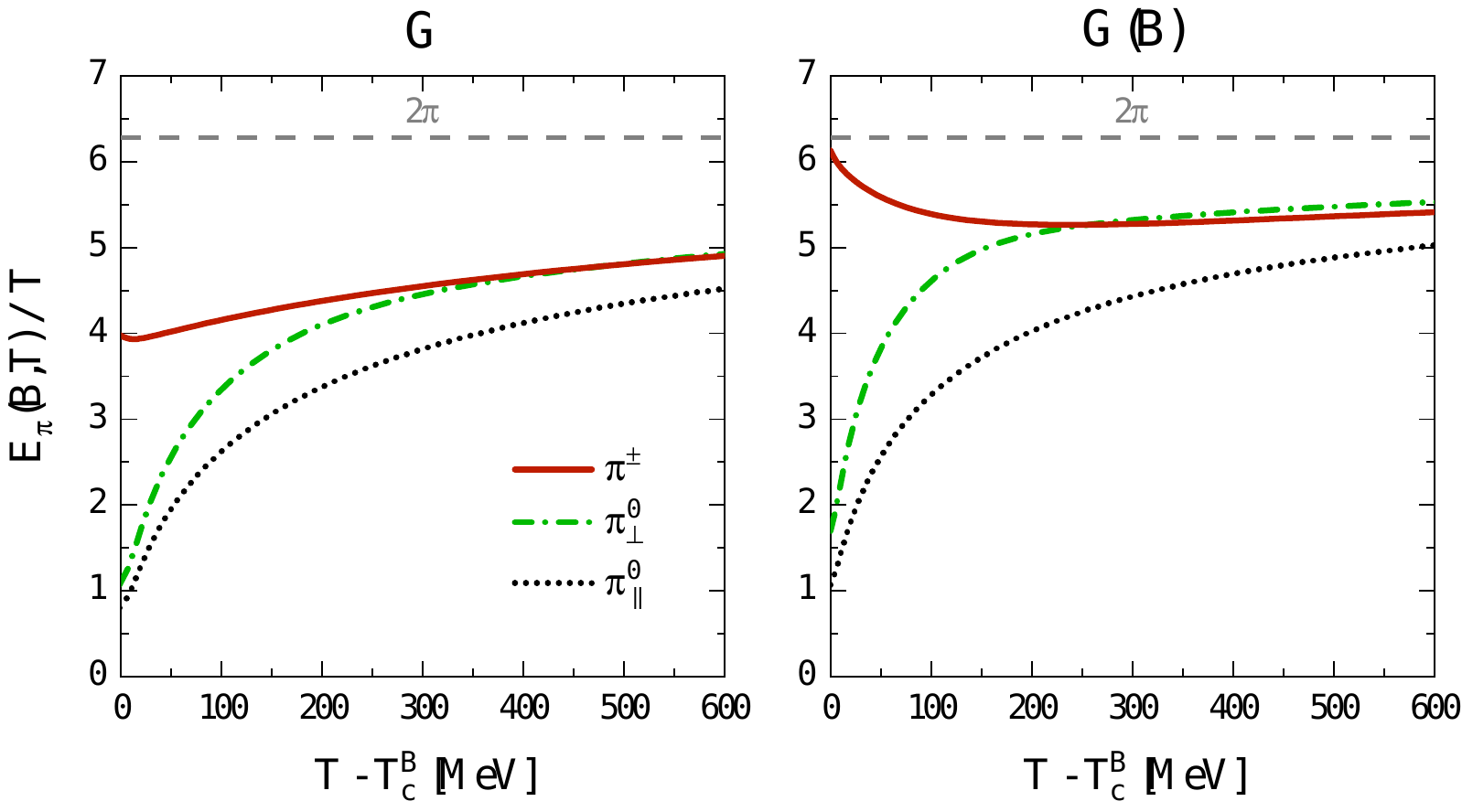}
\caption{Screening masses (energies) of neutral (charged) pions normalized by temperature as functions of the difference between temperature and the pseudocritical temperature at $eB=0.8$~GeV$^2$, for both constant (left) and $B$-dependent (right) coupling constants.}
\label{fig:pi-T_B}
\end{figure}

Regarding charged mesons, the high temperature behavior of their charged screening energies is very similar for pions and kaons, as can be expected from Fig.~\ref{fig:E_fixB}. 
The reasoning leading to Eq.~\eqref{E2q_LES} predicts that their asymptotic limit should be the same as the neutral mesons one, since they are both composed of non-interacting quarks.
Our statement is validated by Fig.~\ref{fig:pi-T_B}, where for the particular pion case it is shown that all masses converge to similar values at high temperature. 
Note that some curves are initially decreasing because the quantity $E_\sl[P](B,T)/T$ artificially diverges at low $T$.


\vspace*{6mm}
\section{Summary \& Conclusions} \label{Conclusions}

In this work we have investigated the screening masses behavior of the pseudoscalar nonet of light mesons at finite magnetic field and temperatures in the context of the SU(3) Nambu-Jona--Lasinio model with the Kobayashi-Maskawa-'t Hooft determinant. 
Through the use of Schwinger's proper time representation of magnetic quark propagators, mesonic actions have been diagonalized by expanding neutral meson fields in the usual Fourier basis, while for charged ones we have applied the Ritus formalism to correctly account for the presence of non-vanishing Schwinger phases with different charges.
Temperature has been included via the Matsubara formalism. 
Since the model is non-renormalizable, ultraviolet divergences have been regularized using the magnetic field independent regularization scheme (see Ref.~\cite{Avancini:2019wed} for advantages of this choice), imposing a 3D sharp cutoff for vacuum contributions.
To incorporate the inverse magnetic catalysis effect in the model, we have allowed for a magnetic-field dependence on one of the model coupling constants $G$, fitted to reproduce LQCD data for the magnetic behavior of the pseudocritical temperature.

While the inclusion of temperature breaks the boost symmetries of the Euclidean group SO(4), differentiating between pole and screening masses, the presence of an external homogeneous magnetic field induces an extra anisotropy further separating neutral screening masses between parallel and perpendicular to the direction of $\vec{B}$. 
As expected from causality, we always get $m_\sl[\mathrm{scr},{\per}] > m_\sl[\mathrm{scr},{\npar}]$ for neutral mesons within our model.
For charged mesons, due to quantization of perpendicular momentum, only a parallel screening energy can be defined, for which we consider its lowest value given by the lowest Landau level.

As for their thermal behavior, at temperatures below $T_c^B$ all screening masses and energies barely change from their $T=0$ value.
Then, around $T\sim T_c^B$, they undergo a transition characterized by a rapid increase, followed by a steady enhancement at high temperatures arising from the medium thermal contribution to the meson energy.
The behavior of the corresponding meson pseudocritical temperature is correlated with the chiral transition one: both increase (decrease) with the magnetic field when a constant ($B$-dependent) coupling is considered.
As a result, at large temperatures screening masses and energies are higher for $G(B)$, since the transition occurred earlier.

Regarding the magnetic behavior of neutral screening masses, a decreasing trend is observed for parallel masses at low temperatures when $G(B)$ is considered, in good agreement with LQCD results at $T=0$~\cite{Ding:2021}.
This behavior is reversed for $T>T_c^B$ resulting in magnetically enhanced parallel masses for $G(B)$, but suppressed for $G$.
This a direct consequence of the magnetic behavior of $T_c^B$, allowing for an earlier (delayed) thermal increase of $m_\sl[\mathrm{scr},{\npar}]$ at higher magnetic fields when $G(B)$ ($G$) is considered.
On the other hand, LQCD simulations~\cite{Ding:2022tqn} show an intricate magnetic behavior, increasing (decreasing) with $eB$ at $T\sim T_c^B$ ($T>T_c^B$).
As discussed in the text, these issues may be influenced by lattice parametrizations and the lack of continuum limit data, as well as uncertainties in the thermomagnetic dependence of $G$ in our NJL model. 
In contrast to longitudinal masses, perpendicular ones are magnetically enhanced at $T=0$.
Moreover, for $G(B)$ they are magnetically enhanced at all temperatures, while for $G$ their behavior is non-monotonic for $T\lesssim 500$~MeV but always increase with $eB$ when $T\gtrsim 500$~MeV.
This means that the magnetic field accelerates the thermal grow of $m_\sl[\mathrm{scr},{\per}]$.
For charged mesons, parallel energies are always magnetically enhanced, independently of $T$.
For  $T\lesssim 200$~MeV, the magnetic energy lies more closely to the point-like case (from above) when $G(B)$ is considered, in better agreement with 
$T=0$ results from LQCD simulations~\cite{Ding:2021} and NJL calculations with vector-axial mixing~\cite{Coppola:2023mmq}.
For $T\gtrsim 200$~MeV, the magnetic enhancement is more pronounced when $G(B)$ is considered.

We have also explored the high-temperature limit of meson masses. 
A straightforward calculation indicates that the energy of two non-interacting quarks at very high temperatures should converge to $E_{2q}\rightarrow 2\pi T$, which we have shown to be fulfilled for both neutral and charged mesons. 
At $B=0$, we have compared different local NJL-type models used in the literature to show how they overestimate the remaining interaction between quarks, leading to pion masses whose increase with temperature is somewhat slower than the one observed in LQCD results~\cite{Bazavov:2019www}. 
We have observed a similar trend at $B \neq 0$ for the screening masses of all pseudoscalar neutral mesons.
Nevertheless, when considering a $B$-dependent coupling, the interaction is reduced at strong magnetic fields, and neutral mesons masses tend more rapidly to the $2\pi T$ limit. 
We have also compared our results for the parallel screening mass of $\pi^0$, considering a field-dependent coupling, with those of LQCD~\cite{Ding:2022tqn}.
In the high-temperature limit, our model results (see also~\cite{Sheng:2021evj}) indicate that the binding decreases with the magnetic field, in contrast with LQCD results.
On the lattice side, this could be related to the lack of continuum limit in the data.
On our NJL model end, there are uncertainties in the thermomagnetic dependence of the coupling of the model, which we have chosen to parameterize independently of temperature.
Moreover, the model can be improved by including mixing with the vector-axial sector, or considering non-local interactions which at $B=0$ already tend to a closer-to-LQCD high-temperature limit.
Lastly, the formalism presented can be extended to consider the pole masses of the studied pseudoscalar mesons, as well as the behavior of their chiral partners, relevant for the analysis of the thermomagnetic dependence of chiral symmetry restoration.


\begin{acknowledgments}
This work has been partially funded by CONICET (Argentina) under Grant No.~PIP 2022-2024 GI-11220210100150CO and ANPCyT (Argentina) under Grant No.~PICT20-01847 (M.C. and N.N.S.), 
by Fundação Carlos Chagas Filho de Amparo à Pesquisa do Estado do Rio de Janeiro (FAPERJ) (Brazil) under Grant No.~SEI-260003/019544/2022 (W.R.T),
by CNPq (Brazil) under Grant No.~308963/2023-7 (S.S.A.), and also is part of the project INCT—FNA (Brazil) under Grant No.~464898/2014-5 (S.S.A. and J.C.S.).
\end{acknowledgments}

\vspace*{6mm}
\appendix

\section{Matrix elements of the inverse meson propagator} \label{app-A}

In this appendix we detail the explicit expressions of the coefficients and polarization functions appearing in the inverse meson propagator of Eq.~\eqref{Gprop}
\begin{align}
{\cal G}_{PP'}(x,x') \:=\: T_{PP'} \ \delta^{(4)}(x-x') - J_{PP'}(x,x') \ .
\end{align}
For the $P,P'=\pi^\pm, K^\pm, K^0, \bar K^0$ subspace, the diagonal matrix elements $T_\sl[P]$ and polarization functions $J_\sl[P](x,x')$ read
\begin{alignat}{6}
T_{\pi^+} &\:=\: T_{\pi^-} &\:=\: \left[ 2 G - K \phi_s \right]^{-1}      \qquad  &; \qquad J_{\pi^+}(x,x') &\:=\: J_{\pi^-}(x',x) &\:=\: c_{ud}(x,x') \ , \\[2mm]
T_{K^+}   &\:=\: T_{K^-}   &\:=\: \left[ 2 G - K \phi_d \right]^{-1}      \qquad  &; \qquad  J_{K^+}(x,x') &\:=\: J_{K^-}(x',x) &\:=\: c_{us}(x,x') \ , \\[2mm]
T_{K^0}   &\:=\: T_{\bar K^0} &\:=\: \left[ 2 G - K \phi_u \right]^{-1}   \qquad  &; \qquad   J_{K^0}(x,x') &\:=\: J_{\bar K^0}(x',x) &\:=\: c_{ds}(x,x') \ ,
\end{alignat}
where the $c_{ff'}(x,x')$ functions are defined in Eq.~\eqref{cff'}.
Note that here and below we use $\phi_f \equiv \phi_f^{B,T}$ for notational shortness.
On the other hand, for the non-diagonal but symmetric matrix elements $T_{PP'}$ of the $P,P'=\pi_3, \eta_0, \eta_8$ subspace we obtain
\begin{align}
T_{\pi_3\pi_3} &\:=\: \dfrac{ K^2 \left( \phi_u + \phi_d \right)^2 - 4 G K \phi_s - 8 G^2}{f} \ , \nonumber\\[3mm]
T_{\eta_0\pi_3} &\:=\: \dfrac{ 2 \left[ K^2 ( \phi_u^{B,T} + \phi_d - \phi_s) - 2 G K \right]
\left( \phi_u - \phi_d \right) }{\sqrt{6} f} \ ,  \nonumber\\[3mm]
T_{\eta_8\pi_3} &\:=\:\dfrac{ \left[ K^2  ( \phi_u + \phi_d + 2 \phi_s) + 4 G K \right]
\left( \phi_u - \phi_d \right) }{\sqrt{3} f} \ ,  \nonumber\\[3mm]
T_{\eta_0\eta_0} &\:=\: \dfrac{K^2 \left[ ( \phi_d - \phi_s)^2 + \phi_u ( \phi_u - 2\phi_d - 2\phi_s) \right]
+ 4 G K ( \phi_u + \phi_d + \phi_s) - 12 G^2 }{3 f/2} \ , \nonumber \\[3mm]
T_{\eta_8\eta_0} &\:=\:  \dfrac{ 2 K^2 \left[ ( \phi_u - \phi_d )^2 + \phi_s ( \phi_u + \phi_d - 2 \phi_s )\right]
- 4 G K \left( \phi_u + \phi_d - 2 \phi_s \right) }{3 \sqrt{2} f} \ ,  \nonumber\\[3mm]
T_{\eta_8\eta_8} &\:=\:  \dfrac{ K^2 \left[ ( \phi_u - \phi_d )^2 + 4 \phi_s ( \phi_u + \phi_d + \phi_s) \right]
- 4 G K \left( 2 \phi_u + 2 \phi_d - \phi_s \right) - 24 G^2 }{ 3 f} \ ,  
\label{eq8}
\end{align}
where
\begin{align}
f \:=\: -4 K^3 \phi_u \phi_d \phi_s + 4 G K^2 \left( \phi_u^2 + \phi_d^2 + \phi_s^2 \right) - 16 G^3 \ .
\end{align}
In turn, the polarization function elements can be expressed as in Eq.~\eqref{Gprop_mix}
\begin{align}
J_{PP'}(x,x') \:=\: \sum_f \gamma^f_{PP'} \ c_{ff}(x,x') \ ,
\end{align}
where the coefficients $\gamma^f_{PP'}$ are given by
\begin{alignat}{8}
\gamma^u_{\pi_3\pi_3} &\:=\: +\gamma^d_{\pi_3\pi_3} &&\:=\: \dfrac{1}{2} \quad &;
\qquad \gamma^s_{\pi_3\pi_3} &\:=\: 0 \quad &;
\qquad \gamma^u_{\eta_0\eta_0} &\:=\: \gamma^d_{\eta_0\eta_0} \:=\: \gamma^s_{\eta_0\eta_0} &&\:=\: \dfrac{1}{3} \ ,
\nonumber \\[3mm]
\gamma^u_{\eta_0\pi_3} &\:=\: -\gamma^d_{\eta_0\pi_3} &&\:=\: \dfrac{1}{\sqrt 6} \quad &;
\qquad \gamma^s_{\eta_0\pi_3} &\:=\: 0 \quad &;
\qquad \gamma^u_{\eta_8\eta_0} &\:=\: \gamma^d_{\eta_8\eta_0} \:=\: -\dfrac{1}{2} \gamma^s_{\eta_8\eta_0} &&\:=\: \dfrac{1}{3\sqrt2} \ ,
\nonumber \\[3mm]
\gamma^u_{\eta_8\pi_3} &\:=\: -\gamma^d_{\eta_8\pi_3} &&\:=\: \dfrac{1}{2\sqrt3} \quad &;
\qquad \gamma^s_{\eta_8\pi_3} &\:=\: 0 \quad &;
\qquad \gamma^u_{\eta_8\eta_8} &\:=\: \gamma^d_{\eta_8\eta_8} \:=\: \dfrac{1}{4}\gamma^s_{\eta_8\eta_8} &&\:=\: \dfrac{1}{6} \ .
\label{gammas}
\end{alignat}


\vspace*{2mm}
\section{Quark propagator and chiral quark condensate}\label{app-B}

In this appendix we detail the calculation of the chiral quark condensate $\phi_f$ defined in Eq.~\eqref{cond_schwPT}.
For its evaluation we need an explicit expression of the quark propagator $\mathcal{S}_f(x,x')$, which can be written in different ways~\cite{Andersen:2014xxa,Miransky:2015ava}.
We adopt the Schwinger proper-time form, given by 
\begin{align}
\mathcal{S}_f(x,x') \:=\: e^{i\Phi_f(x,x')}\,\int \dfrac{d^4p}{(2\pi)^4} \; e^{i p\, (x-x')}\, \mathcal{S}_f(p_{\per},p_{\npar}) \ ,
\label{sfx}
\end{align}
where $\Phi_f(x,x')= Q_f B (x_1+x_1')(x_2-x_2')/2$ is the so-called Schwinger phase, which breaks translational invariance. 
We express the momentum propagator in the Schwinger form~\cite{Andersen:2014xxa,Miransky:2015ava}
\begin{align}
\mathcal{S}_f(p_{\per},p_{\npar}) \:=\: & \int_0^\infty \!d\tau\,
\exp\left[-\tau \left( M_f^2 + p_{\npar}^2 +
\dfrac{\tanh(\tau B_f)}{\tau B_f}\; p_{\per}^2\ - i \epsilon \right) \right] \times \nonumber\\[2mm]
& \left\{\left(M_f-p_{\npar} \cdot \gamma_{\npar} \right)
\, \left[1+i s_f \,\gamma_1 \gamma_2\, \tanh(\tau B_f)\right] -
\dfrac{p_{\per} \cdot \gamma_{\per}}{\cosh^2(\tau B_f)} \right\} \ ,
\label{sfp_schw}
\end{align}
where $p_{\per}=(p_1,p_2)$, $p_{\npar}=(p_3,p_4)$, $\gamma_{\per}=(\gamma_1,\gamma_2)$, $\gamma_{\npar}=(\gamma_3,\gamma_4)$,
$s_f=\mathrm{sign} (Q_f B)$ and $B_f=|Q_fB|$. 
We adopt here the convention $\{\gamma_\mu,\gamma_\nu\}=-2 \delta_{\mu\nu}$. 
The limit $\epsilon\rightarrow 0$ has to be taken at the end of the calculation.
The Dirac space trace in Eq.~\eqref{cond_schwPT} can be straightforwardly calculated now,  leading to the following non-regularized expression for the condensate
\begin{align}
\phi_f \:=\: 4 M_f N_c \int \dfrac{d^4p}{(2\pi)^4} \  \int_0^\infty \!d\tau\, \exp\left[-\tau \left( M_f^2 + p_{\npar}^2 +
\dfrac{\tanh(\tau B_f)}{\tau B_f}\; p_{\per}^2\ - i \epsilon \right) \right] \ .
\end{align}
The Gaussian momenta integrals can be trivially performed, resulting in
\begin{align}
\phi_f \:=\: \dfrac{ N_c M_f}{(2\pi)^2} \int_0^\infty \dfrac{d\tau}{\tau^2} \, e^{-\tau M_f^2} \;  \tau B_f \coth (\tau B_f ) \ .
\label{cond_schwPT-ap}
\end{align}
The remaining integral is divergent and has to be properly regularized. 
We will use the MFIR scheme~\cite{Allen:2015paa,Avancini:2019wed}, where one subtracts from the unregulated integral the $B = 0$ limit and then adds it in a regulated form. 
Extending to finite temperature through the replacements defined in Eq.~\eqref{Trep}, we get, for each flavor, the thermomagnetic chiral condensates
\begin{align}
\phi_f^{B,T} \:=\: \phi_f^{0,T} + \phi_f^{mag,T} \ , \qquad
\begin{cases}
\phi_f^{0,T}   \:\equiv\: -N_c \, M_f \, I_{1f}^{0,T} \\[1mm]
\phi_f^{mag,T} \:\equiv\: -N_c \, M_f \, I_{1f}^{mag,T}
\end{cases} \ .
\label{phif}
\end{align}
The expression of $I_{1f}^{0,T}$ for the 3D cutoff regularization scheme we use in this work can be found in Eq.~\eqref{I1freg}, while the thermomagnetic contribution reads
\begin{align}
I_{1f}^{mag,T} \:=\: \dfrac{T}{2\pi^{3/2}}  \sum_{\ell=-\infty}^{\infty} \int_0^\infty
\dfrac{d z}{z^{3/2}} \ e^{- z ( M_f^2 + \omega_\ell^2)} \ \left[ z B_f  \coth (z B_f) -1 \right] \ .
\end{align}
Interestingly, using the Poisson summation formula $I_{1f}^{mag,T}$ can also be separated as
\begin{align}
I_{1f}^{mag,T} \:=\: I_{1f}^{mag,0} + I_{1f}^{mag,ther} \ ,
\end{align}
where 
\begin{align}
I_{1f}^{mag,0} \:=\: \dfrac{B_f}{2\pi^2} \left[ \ln \Gamma(x_f) - \left(x_f - \dfrac{1}{2}\right) \ln x_f + x_f - \dfrac{\ln{2\pi}}{2} \right] \ , 
\label{i1}
\end{align}
with $x_f=M_f^2/(2B_f)$ and
\begin{align}
I_{1f}^{mag,ther} \:=\: \dfrac{1}{2\pi^2}  \sum_{m=1}^\infty (-1)^m
\int_0^\infty \dfrac{d z }{z^2}\ \exp\left[- \left( z M_f^2 +  \dfrac{m^2}{4z T^2}\right) \right] \  \left[z B_f \coth(zB_f) -1\right] \ .
\end{align}


\vspace*{2mm}
\section{The \texorpdfstring{$B=0$}{t} contribution to the meson polarization function at finite \texorpdfstring{$T$}{t}} \label{app-C}

The polarization function $c^{0,T}_{ff'}$ derived from Eqs.~\eqref{cff'} and~\eqref{cff'_red} can be calculated at $B=0$ using standard techniques, as discussed in~\cite{Ishii:2013kaa}.
Setting the meson Matsubara mode to $\nu_n=0$, the final expression is given by
\begin{align}
c^{0,T}_{ff'}(\vec q\,^2) \:=\: 2 N_c \left\{ \dfrac{ I^{0,T}_{1f} + I^{0,T}_{1f'}}{2} + \left[ \vec q\,^2 + (M_f - M_{f'})^2 \right]  I^{0,T}_{2ff'}(\vec q\,^2) \right\} \ .
\label{cff'reg}  
\end{align}
By performing the sum over Matsubara modes first, the thermal function $I^{0,T}_{1f}$ can be further separated into
\begin{align}
I^{0,T}_{1f} \:=\: I^{vac \vphantom{p}}_{1f} + I^{0,ther}_{1f} \ , 
\label{I1freg}
\end{align}
where for a 3D cutoff regularization, the $B=T=0$ vacuum contribution is given by
\begin{align}
I^{vac \vphantom{p}}_{1f} \:=\: \dfrac{1}{2 \pi^2} \left[ \Lambda\, \sqrt{\Lambda^2 + M_f^2} + M_f^2 \ln\left( \dfrac{M_f}{\Lambda\, + \sqrt{\Lambda^2 + M_f^2}} \right) \right] \ ,
\end{align}
while the thermal one reads
\begin{align}
I^{0,ther}_{1f} \:=\: -\dfrac{2}{\pi^2} \int_0^\Lambda dp\ p^2 \dfrac{1}{E_f} \dfrac{1}{1+\exp(E_f/T)} \ ,
\label{i1temp}
\end{align}
with $E_f^2 = p^2 + M_f^2$. 
Note that although the latter thermal integral is convergent, we have chosen to explicitly regularize it as explained in the main text.
On the other hand, for $\nu_n=0$
\begin{align}
I^{0,T}_{2ff'}(\vec q\,^2) \:=\:  -\dfrac{T}{\pi^2} \, \sum_{\ell=-\infty}^{\infty} \, \int_{0}^1 \, dy  \, \int_0^\Lambda \, dp 
\left[ \dfrac{p}{ p^2+z_\ell(x,\vec q\,^2) } \right]^2 \ , 
\label{I20T}
\end{align}
where
\begin{align}
z_\ell(x,\vec q\,^2) \:=\:  y M_f^2+ (1-y) M_{f'}^2 + \omega_\ell^2 + y(1-y) \, \vec q\,^2 \ .
\end{align}
One could, in principle, perform the sum over Matsubara modes first and separate $I_{2ff'}^{0,T}$ in vacuum plus thermal contributions, as we have done for $I^{0,T}_{1f}$.
Instead, for better numerical convergence we will perform the integral over momentum first, following the prescription of Ref.~\cite{Ishii:2013kaa}.
In this case, the function~\eqref{I20T} becomes complex when the denominator has poles. 
Assuming $z_\ell>0$, i.e. $|M_{f\ell} - M_{f'\ell}|<m_\sl[\rm scr]<M_{f\ell} + M_{f'\ell}$ where $M_{f\ell}=\sqrt{M_f^2+\omega_\ell^2}$, one gets
\begin{align}
I^{0,T}_{2ff'}(\vec q\,^2) \:=\: \dfrac{T}{2 \pi^2} \, \sum_{\ell=-\infty}^{\infty} \, \int_{0}^1 \, dy 
\left[ \dfrac{\Lambda}{ \Lambda^2 + z_\ell(x,\vec q\,^2) } -  \dfrac{\mbox{arccot}\left( \sqrt{ z_\ell(x,\vec q\,^2)}\,/\Lambda \right)}{\sqrt{ z_\ell(x,\vec q\,^2)}}  \right] \ .
\end{align}
We remark that this expression is valid only if the meson mass is below the thermal $\bar q q$ threshold $M_{f\ell}+ M_{f'\ell}$, given by the thermal energy of two non-interacting quarks, as is expected.


\vspace*{2mm}
\section{ The \texorpdfstring{$B\ne 0$}{t} contribution to the meson polarization function at finite \texorpdfstring{$T$}{t} } \label{app-D}

In this appendix we provide details for the calculation of the transformed reduced polarization functions of Eq.~\eqref{cff'_red}
\begin{align}
\mathcal{C}_{ff'}(\bar q,\bar q') \:=\:  
\int d^4x' d^4x \ \mathcal{F}_Q(x,\bar q)^\ast \: c_{ff'}(x,x') \: \mathcal{F}_Q(x',\bar q') \ ,
\label{cff'_barq}
\end{align}
which diagonalize the meson action of Eq.~\eqref{Gdiag} allowing for the calculation of meson masses.
We start at $T=0$ and perform the extension to finite temperature later.
The functions $\mathcal{F}_Q (x,\bar q)$ are solutions of the meson field equations in the presence of an external homogeneous magnetic field, with associated quantum numbers specified by $\bar q$.
For neutral mesons, $\bar q = q =(q^0,\vec q\,)$ stands for the usual four-momentum, with 
${\cal F}_\sl[Q](x,\bar{q})=\exp(iq x)$. 
For charged mesons, ${\cal F}_\sl[Q](x,\bar{q})$ is a gauge dependent Ritus-like function; in this case 
$\bar{q}=(k,\chi,q_3,q_4)$, where $k$ is an non-negative integer related with the so-called Landau level, and the quantum number $\chi$ can be chosen according to the gauge fixing.
Details on this issue can be found in Ref.~\cite{Dumm:2023owj}, where the explicit form of ${\cal F}_\sl[Q](x,\bar{q})$ for three standard gauges is given.

Inserting the quark propagator in the Schwinger proper-time form given in Eq.~\eqref{sfx} we have
\begin{align}
c_{ff'}(x,x') & \:=\: 2 N_c \ \trD \left[ \mathcal{S}_f(x,x') \ \gamma_5 \  \mathcal{S}_{f'}(x',x) \ \gamma_5 \right] \nonumber\\[2mm]
& \:=\: e^{i\Phi_{P}(x,x')}\int \dfrac{d^4v}{(2\pi)^4} \; e^{iv(x-x')} \, c_{ff'}(v_{\per}^2,v_{\npar}) \ ,
\label{cff'_xschw}
\end{align}
where $\Phi_\sl[P](x,x')=\Phi_{Q_{f}}(x,x')+\Phi_{Q_{f'}}(x',x)$ and
\begin{align}
c_{ff'}(v_{\per}^2,v_{\npar}) \:=\: 2 N_c \int \dfrac{d^4p}{(2\pi)^4} \;
\trD \left[\mathcal{S}_{f}(p_{\per},p_{\npar})\,
\gamma_5\, \mathcal{S}_{f'}(p_{\per}-v_{\per},p_{\npar}-v_{\npar}) \, \gamma_5 \right]\ .
\label{cff'_v}
\end{align}
Note that the changes of variables performed in the quark momenta have been done this way in order to lead to ``fermionic'' Matsubara frequencies at finite temperature.
Although not explicitly evident, for the perpendicular momenta this function turns out to depend only on its squared value $v_{\per}^2$, as expected from invariance under rotations along the 3-axis (i.e., the $\vec{B}$-axis).

Replacing Eq.~\eqref{cff'_xschw} into Eq.~\eqref{cff'_barq} we find
\begin{align}
\mathcal{C}_{ff'}(\bar{q},\bar{q}\,')  \:=\: \int \dfrac{d^4v}{(2\pi)^4} \ c_{ff'}(v_{\per}^2,v_{\npar}^2) \ h_\sl[Q](\bar{q},\bar{q}\,'\!, v) \ ,
\end{align}
where $h_\sl[Q](\bar{q},\bar{q}\,'\!,v)$ is a gauge-invariant quantity given by
\begin{align}
h_\sl[Q](\bar{q},\bar{q}\,'\!,v) \:=\:
\int d^{4}x\,d^{4}x' \; {\cal F}_\sl[Q](x,\bar{q})^{\,\ast}\,{\cal F}_\sl[Q](x',\bar{q}\,'\!)\,e^{i\Phi_\sv[Q](x,x')}\,e^{iv(x-x')}\ .
\label{hPiInt}
\end{align}

It can be seen that while $\Phi_\sl[P]$ vanishes in the case of neutral mesons, it does not for charged mesons.
In the first case, it is easy to show that 
$h_{\sl[Q=0]}(q,q',v) = (2\pi)^4\, \delta^{(4)}(q+q')\,  (2\pi)^4 \, \delta^{(4)} (q-v)$, proving that the neutral polarization function becomes diagonal when transformed into ``Fourier space''
\begin{align}
\mathcal{C}_{ff'}^{\sv[\rm neut]}(q,q') & \:=\: (2\pi)^4 \, \delta^{(4)} (q+q') \ c_{ff'}(q_{\per}^2,q_{\npar}^2)\ .
\end{align}
In turn, for charged mesons it has been shown~\cite{Dumm:2023owj} that for three standard gauges one has
\begin{align}
h_{\sl[Q_P]}(\bar{q},\bar{q}\,'\!,v) \:=\:
\delta_{\chi\chi'}\left(2\pi\right)^{4}\,\delta^{\left(2\right)}(q_{\npar}+q_{\npar}^{\prime})
\,\left(2\pi\right)^{2}\,\delta^{\left(2\right)}(q_{\npar}-v_{\npar})\,f_{kk\,'}(v_{\per})\ ,
\label{hPiGauge}
\end{align}
where
\begin{align}
f_{kk\,'}(v_{\per}) \:=\:
\dfrac{4\pi(-i)^{k+k'}}{B_\sl[P]}\,\sqrt{\dfrac{k!}{k^{\prime}!}}\,
\left(\dfrac{2\,v_{\per}^{\;2}}{B_\sl[P]}\right)^{\frac{k'-k}{2}}
L_{k}^{k^{\prime}-k}\Big(\dfrac{2\,v_{\per}^{\;2}}{B_\sl[P]}\Big)\;
e^{-v_{\!\sv[\perp]}^{2}/B_\sl[P]}\;e^{is(k-k')\varphi_{\!\sv[\perp]}}\ .
\label{fkkp}
\end{align}
Here $B_\sl[P]=|Q_\sl[P] B|$ and $s=\mathrm{sign}(Q_\sl[P] B)$, with $Q_\sl[P]=Q_f-Q_{f'}$.
Moreover, $L_k^m(x)$ are generalized Laguerre polynomials, and  $v_{\per} = \tilde v_{\per} (\cos \varphi_{\per}, \sin \varphi_{\per})$. 
Lastly, integrating over $\varphi_{\per}$ one gets
\begin{align}
\mathcal{C}_{ff'}^{\sv[\rm char]}(\bar q,\bar q') & \:=\: (2\pi)^4 \, \delta_{kk'} \: \delta_{\chi\chi'} \: 
\delta^{(2)} (q_{\npar}+q_{\npar}') \ c_{ff'}(k,q_{\npar}) \nonumber\\[2mm] 
c_{ff'}(k,q_{\npar}) & \:=\: \int_0^\infty d\tilde v_{\per} \, \tilde v_{\per} \ c_{ff'}(\tilde v_{\per}^2,v_{\npar}) \ \rho_k(\tilde v_{\per}^{\,2})\ ,
\end{align}
where $\rho_k(\tilde v_{\per}^{\,2}) = 2 (-1)^k\, e^{-\tilde v_{\!\sv[\perp]}^{\,2}/B_\sv[P]} \, L_k(2 \tilde v_{\per}^{\,2}/B_\sl[P])/B_\sl[P]\,$ resembles a normalized distribution function for the perpendicular momenta.
Thus, for charged mesons the polarization function becomes diagonal when transformed into ``Ritus space''. 
In this case, even for $k=0$ one cannot set $v_{\per}=0$, i.e. a charged meson cannot be taken to be at rest.

To obtain results at finite temperature, the replacements detailed in Eq.~\eqref{Trep} have to be performed in the $c_{ff'}(v_{\per}^2,v_{\npar})$ function of Eq.~\eqref{cff'_v}.
Since we are interested in the determination of the screening masses $m_{\rm scr}$ associated with the meson lowest Matsubara mode $n=0$, we will set $\nu_n=0$ in what follows. 
The resulting reduced polarization functions are divergent and thus require to be regularized.
Following the MFIR scheme~\cite{Avancini:2021pmi}, we add an subtract the $B=0$ contribution.
For neutral mesons this leads to
\begin{align}
c^{B,T}_{ff'}(q_{\per}^2, q_3^2) \:=\: 
c^{0,T}_{ff'}(\vec q^{\, 2} = q_{\per}^2 + q_3^2) + c^{mag,T}_{ff'}(q_{\per}^2, q_3^2) \ ,
\label{cff'neutro}
\end{align}
where the first term in the right-hand side corresponds to the regularized $B=0$ contribution, which is implicitly $B$-dependent through the constituent masses $M_f$. 
Its expression is given in Eq.~\eqref{cff'reg} for the choice of a 3D cutoff. 
On the other hand, the thermomagnetic function is given by a trivial generalization of the $T=0$ expression found in Ref.~\cite{Avancini:2021pmi}
\begin{align}
c^{mag,T}_{ff'}(q^2_{\per},q_3^2) \:=\: & \dfrac{N_c\, T}{\pi^{3/2}}  \int_0^\infty dz\ z^{1/2} \int_{0}^1 dy \sum_{\ell=-\infty}^{\infty} \times 
\nonumber \\[2mm]
& \exp\left\{ -z \left[ y M_f^2 + (1-y) M_{f'}^2 + \omega_\ell^2 + y(1-y) (q_{\per}^2 + q_3^2) \right] \right\} \times 
\nonumber \\[2mm]
& \Bigg\{ \exp\left[ -z \left( \dfrac{\gamma_f}{z B_f} - y(1-y)\right) q_{\per}^2 \right] \times 
\nonumber \\[2mm]
& \left[ \left( M_f M_{f'} + \omega_\ell^2 + \dfrac{1}{2 z} - y(1-y) q_3^2 \right) \dfrac{B_f}{\tanh(z B_f)}
+ \left( 1 - \dfrac{\gamma_f}{B_f} q_{\per}^2 \right) \dfrac{B^2_f}{\sinh^2(z B_f)} \right] -
\nonumber \\[2mm]
& \dfrac{1}{z} \left[ M_f M_{f'} + \omega_\ell^2 + \dfrac{3}{2 z} - y(1-y) (q_{\per}^2 + q_3^2) \right] \Bigg\} \ ,
\end{align}
where we have assumed $Q_f=Q_{f'}$ and defined the function
\begin{align}
\gamma_f \:=\: \dfrac{ \sinh(y z B_f) \sinh[(1-y) z B_f]}{ \sinh(z B_f)} \ .
\end{align}

On the other hand, for charged mesons we have
\begin{align}
c_{ff'}^{B,T}(k,\Pi^2) \:=\: c^{0,T}_{ff'}(\vec q\,^2=\Pi^2) + c_{ff'}^{mag,T}(k,\Pi^2) \ ,
\label{cff'carg}
\end{align}
where we have defined $\Pi^2=q_3^2 + (2k+1) B_\sl[P]$, since $\nu_n=0$.
We recall that the $B=0$ expression, regularized through a 3D cutoff, is given in Eq.~\eqref{cff'reg}, while the thermomagnetic contribution reads
\begin{align}
c^{mag,T}_{ff'}(k,\Pi^2) \: = \  & \dfrac{N_c\, T}{\pi^{3/2}} \int_0^\infty dz\ z^{1/2} \int_{0}^1 dy \sum_{\ell=-\infty}^{\infty} \;
e^{ -z \left[ y M_f^2 + (1-y) M_{f'}^2 + \omega_\ell^2 + y(1-y) \Pi^2 \right]} \, \times
\nonumber \\[2mm]
& \Bigg\{  \left[ M_f M_{f'} + \dfrac{1}{2 z} - y(1-y) (\Pi^2-(2k+1)B_\sl[P]) + \omega_\ell^2 \right] \times
\nonumber \\[2mm]
& \hphantom{\Bigg\{} \left[ \dfrac{(1+s_f s_{f'} t_f t_{f'})}{\alpha_+} \left( \dfrac{\alpha_-}{\alpha_+} \right)^k e^{z y(1-y) (2k+1)B_\sv[P]} - \dfrac{1}{z} \right]  \, + 
\nonumber \\[2mm]
& \hphantom{\Bigg\{} 
\dfrac{(1-t_f^2)(1-t_{f'}^2)}{\alpha_+ \, \alpha_-} \left( \dfrac{\alpha_-}{\alpha_+} \right)^k  \left[ \alpha_- + k (\alpha_- - \alpha_+) \right] e^{z y(1-y) (2k+1)B_\sv[P]} \, -
\nonumber \\[2mm]
& \hphantom{\Bigg\{} \dfrac{1}{z} \left[ \dfrac{1}{z}  - \dfrac{1-x^2}{4} (2k+1)B_\sl[P] \right]  \Bigg\} \ ,
\end{align}
where we have introduced the definitions $t_f=\tanh(yzB_f)$, $t_{f'}=\tanh[(1-y)z B_{f'}]$ and $\alpha_\pm=(B_{f'}t_f+B_f t_{f'} \pm B_\sl[P] t_f t_{f'})/(B_f B_{f'})$.

\bibliography{SU3ThermoMesons_bib}

\end{document}